\documentclass[a4paper,11pt]{article}
\usepackage{manuscript}
\title{Generalized hydraulic neural operators turbocharging sewer digital twins}
\author[a]{Ruozhou LIN}
\author[a]{Shixun LI}
\author[a]{Tao SHI}
\author[b]{Lihan HU}
\author[a,c]{Zhiyu ZHANG}
\author[a,d]{
  \authorcr %
  Wenchong TIAN\thanks{
    Co-corresponding author.
    Email: \href{mailto:Wenchong.T@cityu.edu.hk}{\nolinkurl{Wenchong.T@cityu.edu.hk}}.
  }
}
\author[a,e]{
  Zhiguo YUAN\thanks{
    Corresponding author.
    Email: \href{mailto:zhigyuan@cityu.edu.hk}{\nolinkurl{zhigyuan@cityu.edu.hk}}.
  }
}
\affil[a]{School of Energy and Environment, City University of Hong Kong, Hong Kong SAR, China.}
\affil[b]{School of Marine Engineering Equipment, Zhejiang Ocean University, Zhejiang, China}
\affil[c]{College of Environmental Science and Engineering, Tongji University, Shanghai, China.}
\affil[d]{City University of Hong Kong Shenzhen Research Institute, Shenzhen, China.}
\affil[e]{State Key Laboratory of Marine Pollution, City University of Hong Kong, Hong Kong SAR, China.}

\date{}
\begin{document}
  \maketitle
  \label{chap:chapter-six}
  \begin{refsection}
    \section*{Highlights}\label{sec:highlights}

\begin{highlights}
  \item Accelerates SVE-based simulations by more than four orders of magnitude
  \item Enables million-fold speedups for hydraulic calibration with analytic gradients
  \item Supports gradient-based optimization using a differentiable surrogate
  \item Generalizes to new sewer systems without the need for retraining
\end{highlights}

    \begin{abstract}
  \label{sec:abstract}
  Hydraulic digital twins are emerging as the next-generation paradigm for operating and managing sewer systems.
  Conventional hydraulic digital twinning relies on running physics-based models (e.g.,
  Saint-Venant equations), which are computationally expensive and often too slow for real-time applications.
  Although surrogates (e.g., neural networks) can be fast, existing approaches are typically non-generalizable, requiring costly retraining for unseen pipes, networks, or flow conditions.
  Here, we develop a fast, generalized and differentiable multi-branch neural operator trained on solutions of the Saint-Venant equations that predicts in-pipe hydraulics.
  The learned operator reproduces complex hydraulic dynamics with high fidelity while achieving thousand- to million-fold speedups for online prediction, data assimilation, and diagnostics.
  Equally importantly, the learned operator generalizes to unseen pipes, networks, and flow conditions without retraining.
  These advances address key computational barriers to real-time sewer digital twins, opening new opportunities for more responsive flood-risk management and protection of urban water environments.
\end{abstract}
    \begin{keywords}
  \phantomsection\label{sec:keywords}
  digital twin; sewer system; neural operator; Saint-Venant equations
\end{keywords}

    \section{Introduction}\label{sec:introduction}

Urban sewer systems worldwide are facing escalating challenges as external pressures intensify.
The increasing frequency and intensity of extreme rainfall events significantly elevate hydraulic loads~\parencite{westraFutureChangesIntensity2014,zhengOpposingLocalPrecipitation2015a}, often exceeding the design capacity of existing systems.
In parallel, rapid urbanization shrinks pervious surfaces and accelerates hydrological responses~\parencite{fletcherConceptsEvolutionUrban2024,liImpactsUrbanizationWatershed2020,wrightHydroclimatologyFlashFlooding2012}.
When hydraulic capacity is exceeded, sewer surcharge and urban flooding can pose significant socio-economic risks~\parencite{fraehrSuperchargingHydrodynamicInundation2023}.
Beyond flooding, combined sewer overflows discharge untreated wastewater and urban pollutants into receiving waters, causing water pollution and ecosystem degradation~\parencite{butlerUrbanDrainage2018}.
These rising challenges demand effective, forward-looking management strategies, for which digital twins are increasingly recognized as a critical enabling solution~\parencite{webberMovingFutureSmart2022,bartosPipedreamInteractiveDigital2021a}.

A digital twin is a dynamic virtual replica of a physical system continuously updated with real-world data~\parencite{fullerDigitalTwinEnabling2020,taoDigitalTwinIndustry2019,battyDigitalTwinsCity2024}, supporting versatile analytical, predictive and operational workflows to enable smart management and decision-making~\parencite{fullerDigitalTwinEnabling2020,taoDigitalTwinIndustry2019,battyDigitalTwinsCity2024}.
In sewer systems, these applications span flow prediction~\parencite{kimStormwaterDigitalTwin2025,liSimpleEfficientMethod2025,zhangGraphNeuralNetworkbased2024}, sensor data assimilation~\parencite{bartosPipedreamInteractiveDigital2021a,kimStormwaterDigitalTwin2025,kimDigitalTwinModel2024}, early warning~\parencite{kimStormwaterDigitalTwin2025,hlalDigitalTwinTechnology2025}, and proactive control~\parencite{ohModelPredictiveControl2023a,tianFloodingOverflowMitigation2022,tianImprovingInterpretabilityDeep2024,fanEfficientUrbanFlood2025}, among many others.
A sewer hydraulic digital twin comprises a hydraulic model that characterizes the hydraulic behavior of sewer networks, and online data streams that serve as model inputs and support continuous model updates through calibration and validation.
Their integration translates network configurations, hydraulic parameters, and monitoring data into spatiotemporal estimates of sewer flows.
The model should also be swiftly updated within the digital twin to reflect changes in the physical sewer network, such as those associated with aging and renewal.

A common choice of hydraulic model is a physics-based simulator that solves the one-dimensional Saint-Venant equations (SVE), derived from the laws of conservation of mass and momentum~\parencite{linLowcostDataefficientOndevice2025,chenAugmentedHLLRiemann2021,battjesUnsteadyFlowOpen2017}.
Such models offer high-fidelity, physically interpretable hydraulic predictions but incur substantial computational costs~\parencite{fraehrSuperchargingHydrodynamicInundation2023,drainageservicesdepartmentSewerageManualEurocodes2013,drainageservicesdepartmentStormwaterDrainageManual2018,liModelingTransientMixed2024,simonePotentialitiesComplexNetwork2022,crowleyInformationtheoreticSensorPlacement2025}.
Because general analytical solutions to SVE are unavailable, SVE is commonly approximated using numerical schemes that require iterative time stepping at fine spatial and temporal resolutions~\parencite{webberMovingFutureSmart2022,drainageservicesdepartmentSewerageManualEurocodes2013,drainageservicesdepartmentStormwaterDrainageManual2018}.
Surrogate models based on machine learning could accelerate flow simulation by several orders of magnitude~\parencite{kimDigitalTwinModel2024,chenAugmentedHLLRiemann2021,tancikFourierFeaturesLet2020,liModelingTransientMixed2024,simonePotentialitiesComplexNetwork2022,crowleyInformationtheoreticSensorPlacement2025,wuEnhancingInflowInfiltration2025}.
However, existing surrogates are typically case- and task-specific, requiring extensive retraining for new systems or whenever system characteristics change (e.g., through aging or renewal)~\parencite{tianImprovingInterpretabilityDeep2024,garzonMachineLearningBasedSurrogate2022,piadehCriticalReviewRealtime2022}.
Thus, the underlying hydraulic model remains a major bottleneck in the implementation of sewer digital twins.

These limitations fundamentally stem from the absence of an explicit analytical representation of the SVE solution operator, which characterizes the entire family of SVE solutions by directly mapping hydraulic parameters and boundary conditions to their corresponding spatiotemporal flow fields.
Direct access to such an operator would enable generalized, computationally efficient hydraulic modelling, addressing a major barrier to sewer digital-twin implementation.
Here we develop a multi-branch neural operator to approximate the SVE solution operator, with easily measurable manhole water levels serving as real-time data streams to form a robust and streamlined digital twin framework.
Once trained, the learned operator delivers rapid flow estimation and generalizes to unseen pipes, networks and flow conditions without retraining, thereby bringing fully realized sewer digital twins within reach.
The neural operator-based digital twin is then demonstrated via real-time hydraulic prediction, roughness calibration, and leakage detection.

    \section{A generic SVE neural operator}\label{sec:a-generic-sve-neural-operator}

\begin{figure}[p]
  \centering
  \includegraphics{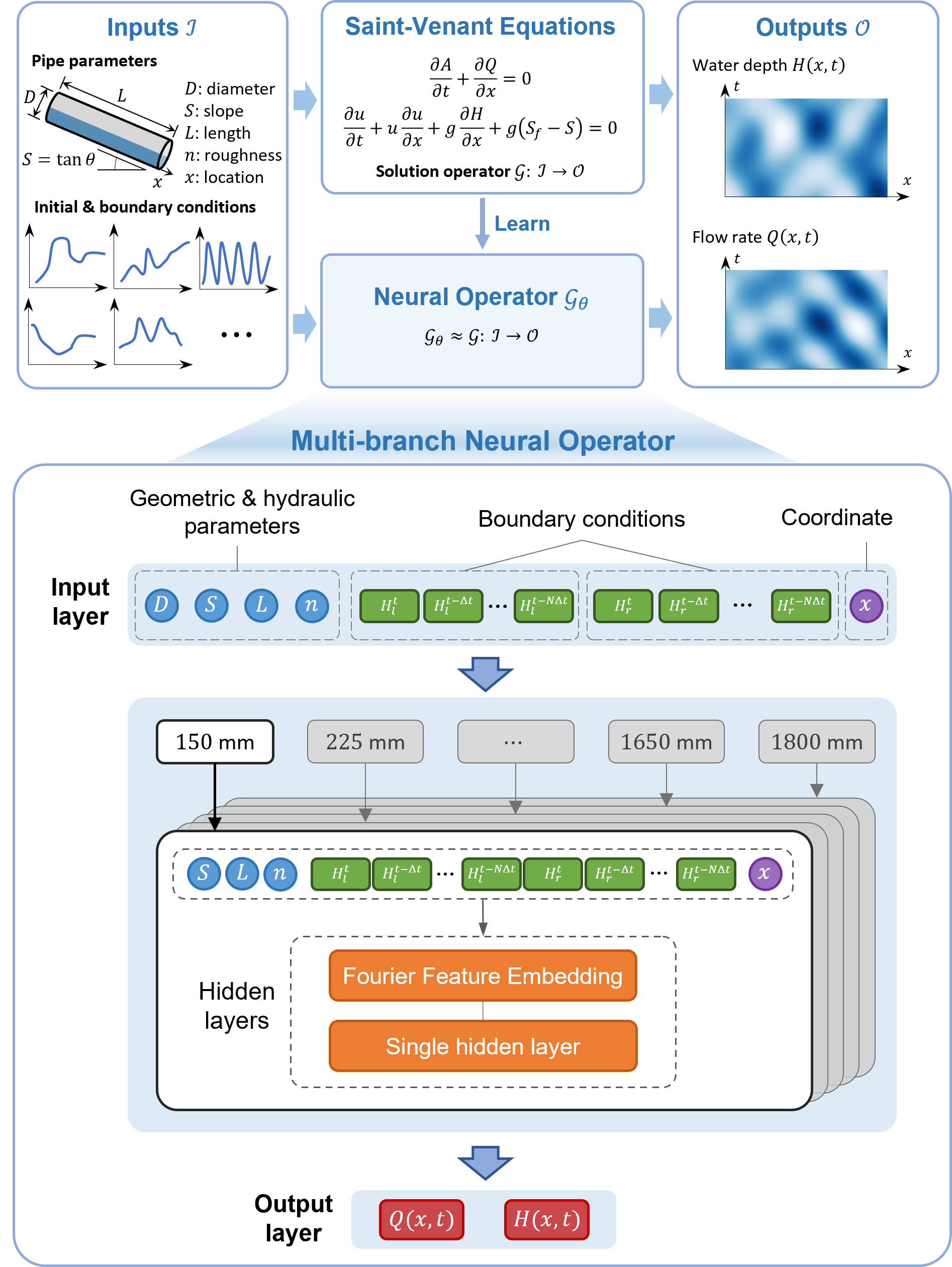}
  \caption{Schematic of the proposed multi-branch neural operator predicting flow rate and water depth at the queried location \(x\) and time \(t\).
  It learns the solution operator of the Saint-Venant equations at the pipe level, directly mapping a pipe's geometric (\(D\), \(S\), \(L\)) and hydraulic (\(n\)) parameters, time series of water depth at the upstream (\(H_{l}^{t},H_{l}^{t-\Delta t},\ldots,H_{l}^{t-N\Delta t}\)) and downstream (\(H_{r}^{t},H_{r}^{t-\Delta t},\ldots,H_{r}^{t-N\Delta t}\)) manholes, and the query coordinate \(x\) to flow rate \(Q(x,t)\) (or equivalently velocity \(u(x,t)\)) and water depth \(H(x,t)\) (or equivalently flow area \(A(x,t)\)).
  The neural operator adopts a multi-branch structure, with each branch corresponding to a specific pipe diameter.
  Given that engineering design codes define only a finite, limited set of standard pipe diameters (typically fewer than 20), diameter is not treated as an input geometric parameter.
  Instead, we train an individual branch for each standard diameter.}
  \label{fig:operator-framework}
\end{figure}

A sewer system is a networked infrastructure composed of pipes and nodes.
In conventional high-fidelity modelling, flow dynamics within each pipe are governed by the SVE, with boundary conditions prescribed at nodes (mainly manholes) located at the two ends of the pipe (Methods, `\hyperref[sec:high-fidelity-physical-modelling]{High-fidelity physical modelling}').
Network-level simulation is achieved by solving SVE for all pipes inside the system simultaneously.
The solution operator of the SVE directly maps the geometric and hydraulic parameters of a pipe---diameter (\(D\)), length (\(L\)), slope (\(S\)), and Manning's roughness coefficient (\(n\))---together with initial and boundary conditions to its corresponding water depth (\(H\)) and flow rate (\(Q\)) solution space.
Because an analytical solution operator is not available, we develop a multi-branch neural operator to approximate it, as illustrated in \autoref{fig:operator-framework}.

Let \(\mathcal{D} \subset\mathbb{R}^{d}\) be a bounded domain.
A standard neural operator \(\mathcal{G}_{\theta}\) seeks to learn the mapping between two infinite-dimensional function spaces:
\begin{equation}
  \label{eq:operator-function-spaces}
  \mathcal{G}_{\theta}: \mathcal{I}\left( \mathcal{D};\mathbb{R}^{d_{I}} \right)
  \rightarrow \mathcal{O}\left( \mathcal{D};\mathbb{R}^{d_{O}} \right)
\end{equation}
where \(\mathcal{I}\) is the space of input functions (e.g., boundary conditions and PDE coefficients); \(\mathcal{O}\) is the space of solution functions; \(d_{I}\) and \(d_{O}\) are the input and output dimensions, respectively.
The proposed operator uses time series of water depth at both ends of the pipe as boundary conditions (\autoref{fig:operator-framework}) because water-level sensors are generally substantially less expensive, more durable, and capable of providing more accurate measurements than flow meters~\parencite{linLowcostDataefficientOndevice2025,linLowcostSoftSensor2025}.
The model and the data constitute a digital twin of the pipe, termed FLOOD (Fast Learned Operator for Online Drainage).
Note that we excluded the initial conditions along the pipe from the input space since they are often unknown in practice and their effects on sewer flows could become negligible after a few minutes~\parencite{linLowcostDataefficientOndevice2025}.
Given the query coordinate (\(x\)), the operator outputs the water depth and flow rate at the specified location and time:
\begin{equation}
  \label{eq:pipe-flow-prediction}
  \left[ \hat{Q}(x,t),\hat{H}(x,t) \right]
  = \mathcal{G}_{\theta}\left( \mathbf{p},\mathbf{b}_{t},x \right)
\end{equation}
where \(\mathbf{p} = [D,S,L,n]^{\top}\) contains the pipe's geometric and hydraulic parameters; 
\(\mathbf{b}_{t} = \left[ H_{l}^{t - N\Delta t},H_{r}^{t - N\Delta t},\ldots,H_{l}^{t},H_{r}^{t} \right]^{\top}\) contains the upstream water depths (\(H(0,\tau)\) or \(H_{l}^{\tau}\) for simplicity) and downstream water depths (\(H(L,\tau)\) or \(H_{r}^{\tau}\) for simplicity) sampled at intervals of \(\Delta t\) over \(\tau \in [t - N\Delta t,t]\) (\ref{sec:si:multi-branch-operator}).
Once trained, such an architecture should enable the operator to generalize to sewer pipes with diverse boundary conditions.

Different from other input variables, the space of diameter comprises a finite set of nominal values with typically fewer than 20 standard options.
Therefore, for dimensionality reduction, we divide the SVE operator into a finite number of branches, 14 in this study according to the Hong Kong design standard, most of which are widely adopted worldwide~\parencite{drainageservicesdepartmentSewerageManualEurocodes2013,drainageservicesdepartmentStormwaterDrainageManual2018}.
Each branch corresponds to only one nominal diameter (e.g., \qty{150}{\milli\metre}) (\autoref{fig:operator-framework}).
During inference, the multi-branch operator deterministically selects the corresponding branch operator based on pipe diameter, while the remaining geometric and hydraulic parameters, boundary conditions, and query coordinate are provided as the input to that branch.
Each branch operator employs a Fourier feature embedding layer~\parencite{tancikFourierFeaturesLet2020,liModelingTransientMixed2024} followed by a single hidden layer to map input to the final output.

\FloatBarrier

    \section{Pipe-level digital twins}\label{sec:pipe-level-digital-twins}

\subsection{Learning across diverse hydraulic regimes}\label{sec:learning-across-diverse-hydraulic-regimes}

Training a generalizable neural operator requires massive amounts of representative, high-fidelity data with high spatiotemporal resolution, spanning both the infrastructure design space and the hydraulic regimes encountered in sewer systems.
We aggregated the training and validation data by solving the Saint-Venant equations using a first-order Godunov scheme with the HLL Riemann solver (Methods, `\hyperref[sec:high-fidelity-physical-modelling]{High-fidelity physical modelling}')~\parencite{zhangGraphNeuralNetworkbased2024,piadehCriticalReviewRealtime2022,shiODEbasedSwiftDynamic2025}.
Diameter, length, slope and roughness were sampled across code-defined engineering ranges with combinations violating design constraints excluded, yielding 2,871 plausible pipe configurations (\ref{sec:si:hydraulic-parameter-combinations}).

Boundary conditions pose a greater challenge because arbitrary upstream and downstream time series form an effectively unbounded space, and these series are not independent but physically coupled in real systems.
We therefore designed a seven-segment computational domain that generated diverse yet physically coupled conditions around a central observation segment (\ref{sec:si:seven-segment-data-generation}).
At the upstream boundary, a zero-slope segment established a controlled subcritical inflow.
Immediately downstream, a transition segment gradually increased the slope to the target value, followed by a buffer segment that attenuated numerical artefacts before flow entered the observation segment.
Downstream of the observation segment, a second buffer segment and a transition segment led to a terminal segment with a higher roughness value that generated flow-dependent hydraulic resistance through pipe-wall friction.
Although the roughness was fixed within each scenario, the resulting resistance was weak under low-flow conditions but increased as incoming flow increased, producing backwater effects and surcharge during high-flow events.
This configuration reduced the downstream boundary-condition space from unrestricted time series to a controlled resistance mechanism that responded dynamically to incoming flow.
We systematically varied the roughness of the terminal segment to expose each pipe to free-surface and pressurized states, backwater effects, transcritical behavior, and transitions between hydraulic regimes (\ref{sec:si:seven-segment-data-generation}).

We hypothesized that, over short time windows, the dominant temporal patterns of sewer inflows could be approximated using sinusoidal functions.
We therefore varied their amplitudes, frequencies, phases and offsets to span the magnitudes and timescales characteristic of steady flows, gradually varying dry-weather flows, wet-weather hydrographs and rapid pump-driven fluctuations (\ref{sec:si:training-boundary-conditions}).
Combining these synthetic inflows with the sampled pipe properties and downstream resistance produced nearly 0.7 million scenarios and more than 3 billion data points.
After preprocessing, approximately 0.9 and 0.2 billion representative data points were retained for training and validation, respectively (\ref{sec:si:data-preprocessing}).

We chose the time series of water depth from \(t - 5\Delta t\) to \(t\), where \(\Delta t = 1\) min, as inputs to the proposed operator based on an analysis of the characteristic curves~\parencite{linLowcostDataefficientOndevice2025}.
The hyperparameters for each branch were determined by grid search, resulting in approximately 280k trainable parameters per branch on average (\ref{sec:si:multi-branch-operator}).

\subsection{High-fidelity and fast prediction without retraining}\label{sec:high-fidelity-and-fast-prediction-without-retraining}

Although the operator was trained exclusively on synthetic data driven by simple sinusoidal inflows, we tested it on approximately 0.4 billion points from real infrastructure, using either field measurements or field-driven simulations, and spanning unseen combinations of geometric and hydraulic parameters and boundary conditions (\ref{sec:si:data-preprocessing} and~\ref{sec:si:test-pipes}).

\begin{figure}[p]
  \centering
  \includegraphics{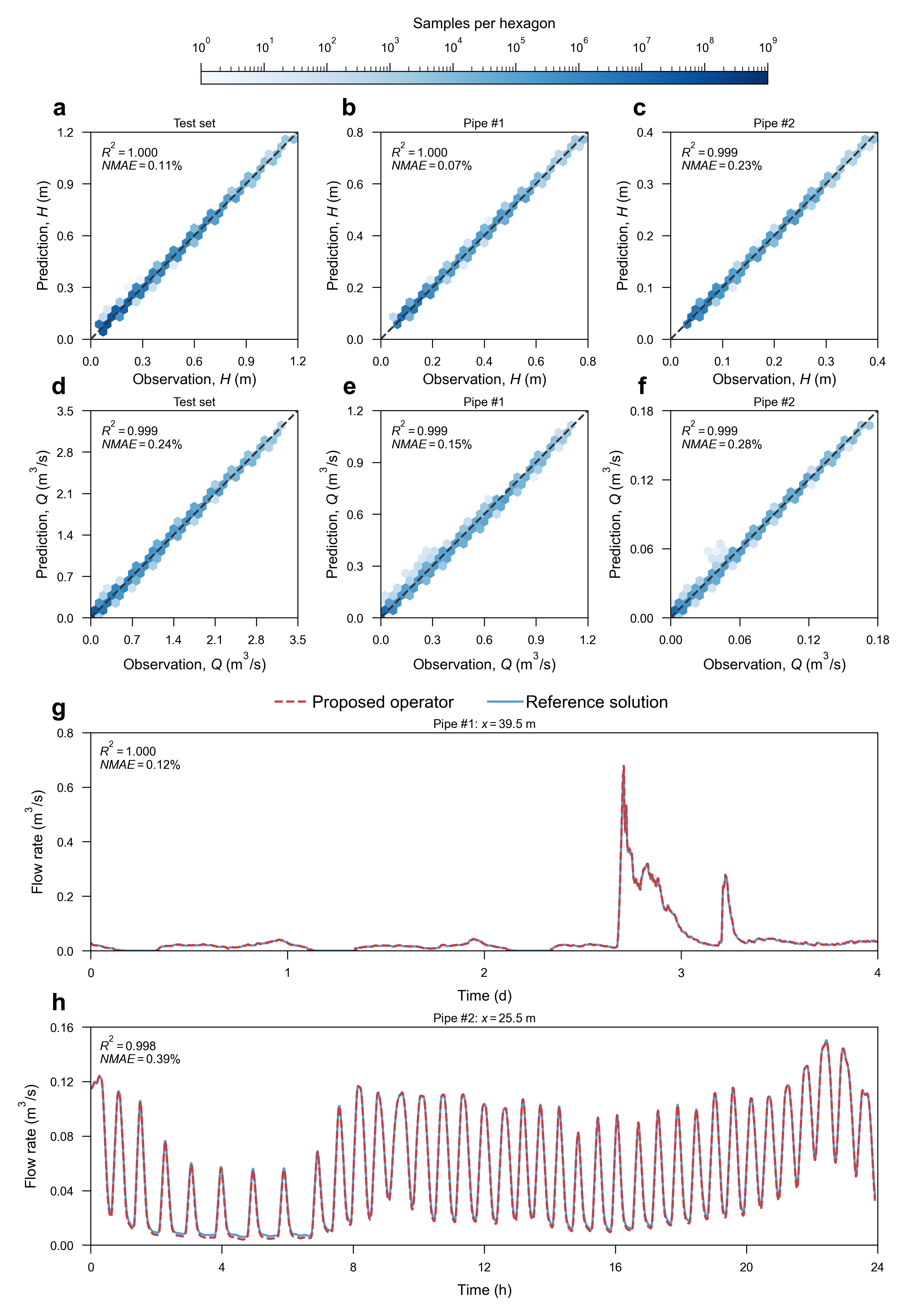}
  \caption{Pipe-level neural operator digital twins well-reproduce in-pipe hydraulic dynamics.
  Comparisons of learned operator predictions with reference solutions for water depth (\(H\)) (a--c), and flow rate (\(Q\)) (d--f), for the test set (a,d), Pipe \#1 (b,e) and Pipe \#2 (c,f), with \(R^2\) and normalized mean absolute errors (NMAE). g,h, Representative flow-rate time series predicted by the learned operator and the reference solution for Pipe \#1 (g) and Pipe \#2 (h) at the indicated locations, with \(R^2\) and NMAE.}
  \label{fig:pipe-level-performance}
\end{figure}

We first evaluated whether the learned SVE neural operator can serve as a generalized pipe-level digital twin that efficiently reproduces high-fidelity hydraulic states.
On the test set, the learned operator achieved \(R^2\) of 1.000 for water depth and 0.999 for flow rate (\autoref{fig:pipe-level-performance}a,d), with normalized mean absolute errors (NMAE, defined in \ref{sec:si:evaluation-metrics}) of 0.11\% and 0.24\%.
We also selected two real-life pipes (called Pipes \#1 and \#2, \autoref{tab:test-pipe-parameters}) with representative scenarios from the test set for demonstration (\autoref{fig:pipe-level-performance}b,c,e,f,g,h).
The learned operator accurately reproduced the spatial distribution and temporal propagation of water depth and flow rate along Pipes \#1 and \#2, including transient variations driven by changing upstream and downstream conditions.
For Pipe \#1, the model achieved \(R^2\) of 1.000 (water depth) and 0.999 (flow rate), with NMAE of 0.07\% and 0.15\%.
For Pipe \#2, \(R^2\) reached 0.999 for both water depth and flow rate, with NMAE of 0.23\% and 0.28\%.
This high level of fidelity was achieved alongside drastically reduced computational overhead.
The operator achieved approximately 5,500- and 3,100-fold speedups for Pipes \#1 and \#2, reducing computation times from 121.0 to \qty{0.022}{\second} and from 251.2 to \qty{0.081}{\second} on a personal laptop, respectively.

The above results also indicate that the proposed solution operator exhibits generalization capability: once trained, the operator can be applied to unseen sewer pipes and flow conditions without retraining, as shown by its performance on pipes whose geometric and hydraulic parameters and boundary conditions were not included in the training or validation sets.
Despite the training data being generated using simple synthetic sinusoidal boundary conditions, the trained operator was tested on scenarios (Pipes \#1 and \#2) simulated based on field measurements, covering dry and wet weather flows, pump-driven high-frequency fluctuations, backwater effects, and severe surcharge conditions.
The strong performance observed across these cases indicates that the proposed method learned a generalized operator for the Saint-Venant equations rather than a case-specific mapping for a limited number of pipes and inflow patterns.
This distinction is critically important for practical application, because retraining a model for every pipe would limit its usefulness for real sewer digital twins.

\subsection{Instant pipe roughness calibration}\label{sec:instant-pipe-roughness-calibration}

Pipe roughness (\(n\)) is a key hydraulic parameter that characterizes the resistance to flow caused by friction along the pipe wall.
In real-world sewer systems, effective pipe roughness can vary over time owing to various factors, such as pipe ageing, sediment deposition and biofilm growth.
Pipe roughness therefore needs to be reassessed periodically.

Conventionally, the roughness (\(n^{*}\)) is calibrated by minimizing the discrepancy between a hydraulic variable (water depth or flow rate) observed by a sensor and that simulated using SVE at the monitored location.
The discrepancy can be quantified by the squared \(L_{2}\) norm.
For example, using the observed flow rate series (\(\mathbf{Q}_{\mathrm{obs}}\)) and simulated flow rate series (\(\mathbf{Q}_{\mathrm{sim}}\left( x_{s},n \right)\)) at the sensor location (\(x_{s}\)), the roughness calibration problem can be formulated as:
\begin{equation}
  \label{eq:roughness-calibration}
  n^{*} = \operatorname*{arg\,min}_{n} J(n),
  \qquad \text{where } J(n)
  = \left\| \mathbf{Q}_{\mathrm{obs}}
  -\mathbf{Q}_{\mathrm{sim}}\left( x_{s},n \right) \right\|_{2}^{2}
\end{equation}

Starting from an initial guess, \(n^{*}\) can be determined by gradient-based methods such as L-BFGS-B, which requires \(\frac{dJ(n)}{dn}\) as the gradient.
Conventional SVE-solver implementations are generally non-differentiable.
They must therefore approximate \(\frac{dJ(n)}{dn}\) through repeated evaluations of \(\mathbf{Q}_{\mathrm{sim}}\left( x_{s},n \right)\).
Moreover, every evaluation requires a time-consuming full-domain hydraulic simulation (\(\Omega = [ 0,L] \times [ 0,T]\)), although only the solution at the sensor location (\(\Gamma_{s} = \left\{ x_{s} \right\} \times [ 0,T]\)) is used for calibration.
These requirements make conventional roughness calibration computationally demanding.

The learned operator can be used for efficient calibration by minimizing the differences between the estimated flow rate (\(\mathbf{Q}_{\mathrm{sim}}\left( x_{s},n \right)\)) and observed flow rate.
In principle, the proposed method should be substantially more computationally efficient for three reasons.
First, the learned operator accelerates hydraulic simulation, as demonstrated by the pipe-level results.
Second, the learned operator directly queries \(\mathbf{Q}_{\mathrm{sim}}\left( x_{s},n \right)\) at the sensor location, while a conventional SVE solver needs to extract \(\mathbf{Q}_{\mathrm{sim}}\left( x_{s},n \right)\) from a full-domain simulation.
Third, its differentiable architecture enables \(\frac{dJ(n)}{dn}\) to be obtained directly and explicitly through automatic differentiation, avoiding repeated model evaluations and the step-size-dependent errors associated with numerical gradient approximation.

We evaluated the proposed operator-based calibration method on two cases not included in the operator training set: Pipes \#3 and \#4 (Methods, `\hyperref[sec:evaluation-cases-and-study-sites]{Evaluation cases and study sites}').
Pipe \#3 is a real-life pipe with field measurements under free-surface flow conditions, and Pipe \#4 is a synthetic pipe under surcharge conditions.
The proposed method yielded Manning's roughness coefficients of 0.025 and 0.015 for Pipes \#3 and \#4, respectively, closely matching the reference values of 0.025 and 0.016.
Compared with the conventional SVE-based calibration workflow, the proposed operator-based workflow reduced the calibration times from \qty{16.4}{\hour} and \qty{235.0}{\hour} to \qty{0.259}{\second} and \qty{0.715}{\second}, corresponding to speedups of 228,284-fold and 1,182,962-fold, respectively.

\FloatBarrier

    \section{Network-level digital twins}\label{sec:network-level-digital-twins}

\subsection{Pipe-to-network generalization without retraining}\label{sec:pipe-to-network-generalization-without-retraining}

The learned operator can be extended to network-level digital twins without retraining.
When water depths at all manholes are monitored, the extension is straightforward.
For a sewer network that has \(N_{u}\) unmonitored manholes, we propose to estimate the unmonitored water depths (\(\hat{\mathbf{H}}^{t} \in \mathbb{R}^{N_{u}}\)) at time \(t\) by solving the following optimization problem:
\begin{equation}
  \label{eq:unmonitored-depth-inference}
  \hat{\mathbf{H}}^{t}
  = \operatorname*{arg\,min}_{\mathbf{H}_{t}} r\left( \mathbf{H}^{t} \right),
  \qquad \text{with } 
  r\left( \mathbf{H}^{t} \right)
  = \sum_{i}\left( \sum_{j \in \mathcal{A}_{i}}\left[ \mathcal{G}_{\theta}\left( \mathbf{p}_{j},\mathbf{b}_{j,t},x_{j} \right) \right]_{1} \right)^{2}
\end{equation}
where \(r\) is the sum of squared mass balance residuals at all unmonitored manholes; \(\mathcal{A}_{i}\) is the set of pipes connected to the \(i\)th unmonitored manhole; \(\mathcal{G}_{\theta}\) denotes the learned operator, with \(\left[ \mathcal{G}_{\theta}\left( \boldsymbol{\cdot} \right) \right]_{1}\) representing the first output variable, i.e., the flow rate; \(\mathbf{p}_{j}\) contains the \(j\)th pipe's geometric and hydraulic parameters; \(\mathbf{b}_{j,t}\) contains the water depths at both ends of pipe \(j\) as boundary conditions, which are either obtained from online monitoring or treated as variables to be estimated during the iterative optimization; \(x_{j}\) denotes the local coordinate of manhole \(i\) with respect to pipe \(j\).

Since the learned operator is differentiable, we can directly calculate \(\frac{\partial r\left( \mathbf{H}^{t} \right)}{\partial \mathbf{H}^{t}}\) via automatic differentiation, enabling gradient-based algorithms to efficiently determine \(\hat{\mathbf{H}}^{t}\).
We used the L-BFGS algorithm with a strong Wolfe line search here.
Once the unmonitored water depths are reconstructed, flows throughout the sewer network can be simulated using the trained operator.

\begin{figure}[p]
  \centering
  \includegraphics{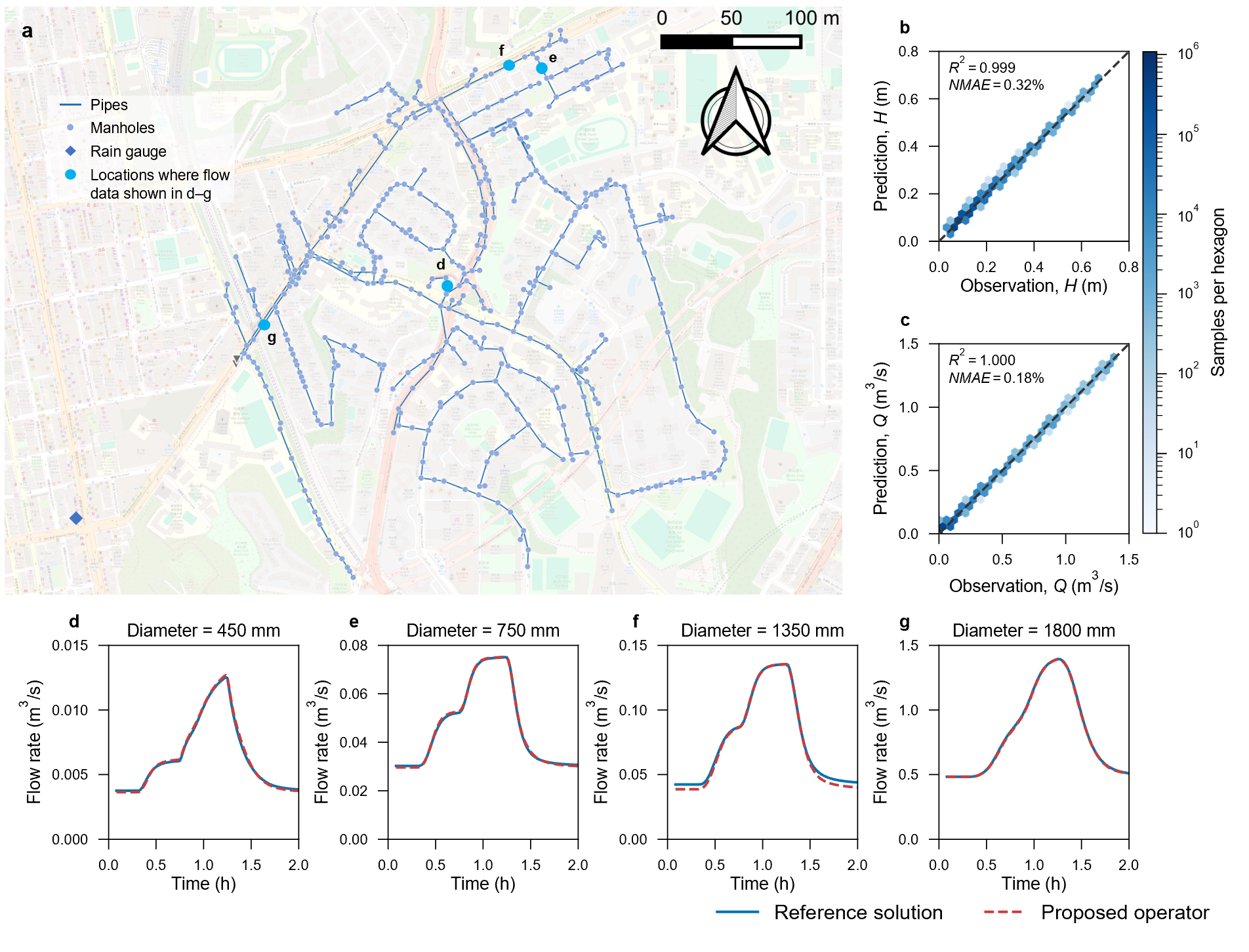}
  \caption{Pipe-to-network generalization performance for network-level simulation. a, Studied sewer network. b,c, Agreement between predictions and reference solutions for water depth (\(H\)) and flow rate (\(Q\)), respectively. d-g, Flow rate hydrographs at selected locations spanning diverse pipe hydraulic parameters and flow conditions.}
  \label{fig:network-depth-inference}
\end{figure}

We evaluated this approach using a sewer network located in Kowloon City District, Hong Kong, covering \qty{1.2}{\kilo\metre\squared} and containing 485 pipes with a total length of \qty{14.44}{\kilo\metre}.
We first considered an idealized fully monitored case, in which water depths were available at all manholes.
The proposed method achieved high fidelity with \(R^2\) values of 0.999 and 1.000 and NMAE values of 0.32\% and 0.18\% for water depth and flow rate, respectively.
The hydrographs further illustrate the close agreement between the proposed operator and the high-fidelity reference (\autoref{fig:network-depth-inference}).
The proposed method reduced simulation time from \qty{7720.9}{\second} for the high-fidelity SVE solver to \qty{0.403}{\second}, achieving a 19,165-fold speedup.

\begin{figure}[p]
  \centering
  \includegraphics{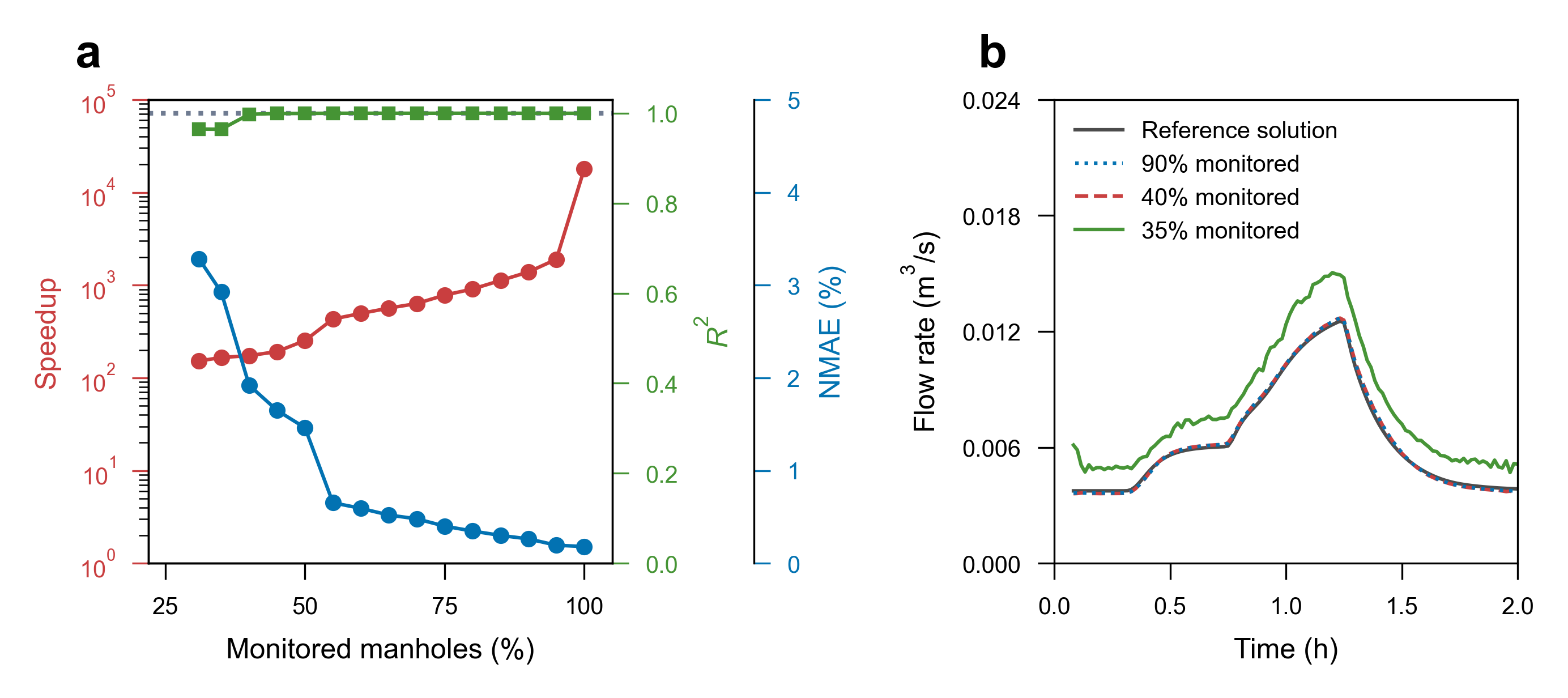}
  \caption{Model performance under different sensor densities (percentage of monitored manholes). a, Prediction accuracy and computational speedup relative to the high-fidelity SVE solver, quantified by coefficient of determination (\(R^2\)) and normalized mean absolute errors (NMAE) for flow rate. b, Flow rate hydrographs at location d (\autoref{fig:network-depth-inference}a) under different sensor densities.}
  \label{fig:network-level-performance}
\end{figure}

We further evaluated partially monitored situations, where water depths at unmonitored manholes were inferred using \autoref{eq:unmonitored-depth-inference}.
This setting is more challenging because the iterative solution introduces additional computational cost.
Also, errors in inferred manhole depths propagate through adjacent pipes.
We tested 14 different scenarios with monitored manholes ranging from 30\% to 95\%.
For each percentage scenario, 50 sensor placements were randomly generated, and the placement with the highest \(R^2\) for flow rate is presented here to focus on the effect of sensor density rather than sensor locations (sensor location optimization is an independent research question~\parencite{simonePotentialitiesComplexNetwork2022,crowleyInformationtheoreticSensorPlacement2025,wuEnhancingInflowInfiltration2025,zhengNovelFrameworkOptimization2025}).
The results reveal a clear trade-off between monitored percentage and performance (\autoref{fig:network-level-performance}).
At 60-95\%, the proposed method could achieve a speedup of more than 500 times while maintaining an \(R^2\) of 1.0 and an NMAE below 0.7\% for flow rate.
At around 40\%, the proposed method still achieved a 174-fold speedup with \(R^2 = 0.998\) and \(\mathrm{NMAE} = 1.9\%\).
At a sensor density of 30\% or lower, the resulting performance (\(R^2 = 0.965\)) is more suitable for applications where moderate fidelity is acceptable, rather than supporting high-fidelity network-level digital twins.
Overall, these results demonstrate the potential of the proposed method, providing a practical route towards real-time sewer digital-twin applications.

\FloatBarrier

    \section{Rapid leakage identification: an application example}\label{sec:rapid-leakage-identification-an-application-example}

Defects in sewer pipes can allow groundwater to infiltrate into the pipes or wastewater to exfiltrate into the surrounding environment.
Infiltration imposes additional hydraulic loads on sewer conveyance and wastewater treatment systems, whereas exfiltration contaminates surrounding soil and groundwater.
Therefore, rapid leakage localization and quantification are important but remain challenging.
Here, we demonstrate the use of the learned operator to rapidly locate leaking pipes and quantify leaked volumes based on water-level measurements at manholes.

In the absence of leakage, the learned operator accurately captures the relationship between hydraulic inputs and outputs, resulting in small mass balance residuals at each manhole.
In contrast, leakage alters this relationship by allowing water to enter or leave the pipe.
Because this unaccounted flow is not represented by the learned operator, the estimated flow rates no longer satisfy node-scale mass balance, resulting in large residuals at the adjoining manholes.
Thus, a pipe is likely to have leakage when both endpoints exhibit high node residuals.
We therefore define a pipe leakage evidence score to locate pipes with a high likelihood of leakage (Methods, `\hyperref[sec:leakage-localization-and-quantification]{Leakage localization and quantification}').
Beyond localization, the learned operator can be used to estimate the infiltration/exfiltration volume based on mass conservation (Methods, `\hyperref[sec:leakage-localization-and-quantification]{Leakage localization and quantification}').

\begin{figure}[p]
  \centering
  \includegraphics{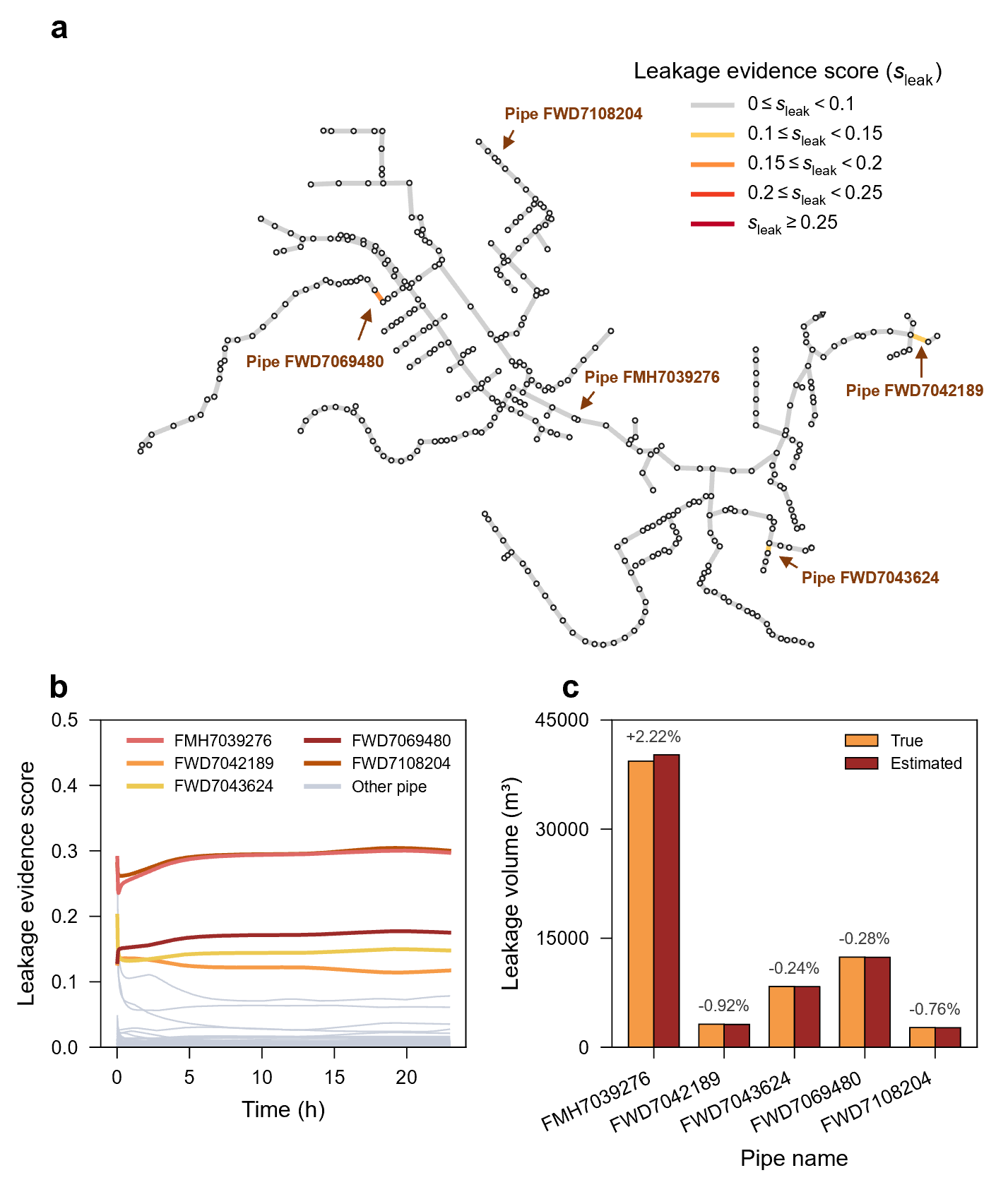}
  \caption{Sewer leakage localization and quantification.
  a, Spatial distribution of pipe leakage evidence scores.
  b, Evolution of pipe leakage evidence scores.
  c, Comparison of true and estimated leakage volumes.}
  \label{fig:leakage-identification}
\end{figure}

We evaluated the method using a sewer network in the Greater Bay Area, China, containing 397 pipes with a total length of \qty{11.17}{\kilo\metre}.
Significant groundwater infiltration has been detected by the local water utility.
For this conceptual demonstration, water depth measurements at all manholes were assumed.
Infiltration was imposed at the midpoints of five randomly selected pipes (\autoref{fig:leakage-identification}a) in a \qty{24}{\hour} simulation.
The five pipes with imposed infiltration occupied the top five positions when ranked by the pipe leakage evidence score (\autoref{fig:leakage-identification}b).
The leaked volumes at all five locations were also successfully quantified with relative errors below 2.5\% (\autoref{fig:leakage-identification}c).
The evaluation took \qty{0.33}{\second}.

Together, these results demonstrate the potential of the learned operator for rapid leakage localization and quantification, its successful generalization to an unseen sewer network, and its high computational efficiency.

\FloatBarrier

    \section{Discussion}\label{sec:discussion}

Sewer digital twins require hydraulic engines that are accurate, computationally efficient, differentiable, and generalized across sewer systems.
This study demonstrates a multi-branch neural operator that meets these requirements by learning the pipe-level solution operator of the Saint-Venant equations.
Without retraining, it accurately reproduces hydraulic dynamics across unseen pipes and flow conditions, and was successfully applied to two real-life sewer networks for hydraulic simulation and leakage localization and quantification.
This combination of efficiency and generalization provides a practical foundation for sewer digital twins.

The exceptionally large speedup arises from three complementary sources.
First, the trained operator directly maps geometric and hydraulic parameters and boundary conditions to the queried hydraulic states through a single neural-network evaluation, avoiding the iterative time stepping required by conventional SVE solvers.
Second, its differentiable architecture allows gradients of objective functions with respect to hydraulic parameters to be obtained directly through automatic differentiation.
Conventional non-differentiable solvers generally require repeated simulations to approximate these gradients numerically.
Third, whereas conventional solvers must calculate the complete spatiotemporal hydraulic domain before extracting results at monitored locations, the proposed operator can be queried only at the coordinates required by a particular application.
Together, these advantages substantially accelerate computation-intensive digital-twin workflows.

The generalization of the proposed operator arises at two levels.
At the pipe level, the operator receives a comprehensive description of the hydraulic problem, including pipe geometry, hydraulic parameters, boundary conditions and query coordinates.
Similar to an SVE solver, it maps these problem specifications to the corresponding hydraulic states rather than learning the response of a fixed pipe, enabling transfer to unseen pipes and flow conditions.
At the network level, the operator learns the fundamental element from which sewer networks are assembled: the pipe.
The same trained operator can therefore be applied to every pipe and coupled through mass conservation, enabling simulation of unseen sewer networks without retraining.
These two levels of generalization were demonstrated across diverse pipe- and network-scale cases.
Applying the same operator to roughness calibration and leakage localization and quantification further demonstrates that its generalization extends beyond forward simulation to inverse and diagnostic tasks.

Several opportunities remain for further development.
First, extending the operator to non-circular conduits, control structures, and sewer-overland interactions would broaden its application to more complex sewer systems.
Second, network-scale digital twins under sparse monitoring could be strengthened by jointly optimizing sensor coverage and placement.
Identifying the most informative sensor configurations could improve hydraulic accuracy while reducing monitoring costs.
Finally, more diverse training data, together with stronger physics-guided architectures, could further improve reliability across a broader range of hydraulic conditions, while evaluation across additional sewer systems would strengthen the evidence for the transferability of the proposed method.

Together, these advances in computational efficiency, differentiability, and generalization of sewer digital twins create new opportunities for more responsive sewer management.

    \section{Methods}\label{sec:methods}

\subsection{High-fidelity physical modelling}\label{sec:high-fidelity-physical-modelling}

In convention, high-fidelity sewer flow modelling uses the 1D Saint-Venant equations (SVE) derived from the conservation laws of mass and momentum to describe the motion of water in channels, rivers, and pipes~\parencite{butlerUrbanDrainage2018,shiODEbasedSwiftDynamic2025}.
Given a circular pipe with known diameter (\(D\)), length (\(L\)), slope (\(S\)), and Manning's roughness coefficient (\(n\)), SVE is expressed as:
\begin{subequations}
  \label{eq:saint-venant}
  \begin{equation}
    \label{eq:mass-conservation}
    \frac{\partial A}{\partial t} + \frac{\partial Q}{\partial x} = 0
  \end{equation}
  \begin{equation}
    \label{eq:momentum-conservation}
    \frac{\partial u}{\partial t}
    + u\frac{\partial u}{\partial x}
    + g\frac{\partial H}{\partial x}
    + g\left( S_{f} - S_{0} \right)
    = 0
  \end{equation}
\end{subequations}
where \(x\) denotes the distance from the pipe inlet (\unit{\metre}); \(t\) is time (\unit{\second}); \(Q(x,t)\), \(H(x,t)\), \(u(x,t)\), and \(A(x,t)\) are, respectively, the flow rate (\unit{\cubic\metre\per\second}), water depth (\unit{\metre}), velocity (\unit{\metre\per\second}) and cross-sectional flow area (\unit{\square\metre}) at distance \(x\) and time \(t\); \(g\) is the gravitational acceleration (\qty{9.81}{\metre\per\second\squared}); \(S_{0}\) and \(S_{f}(x,t)\) are the pipe slope and the frictional slope, respectively; \(S_{f}\) can be calculated using Manning's formula:
\begin{equation}
  \label{eq:manning-friction-slope}
  S_{f} = \frac{n^{2}Q|Q|}{A^{2}R^{\frac{4}{3}}}
\end{equation}
where \(R(x,t)\) is the hydraulic radius (\unit{\metre}).

We used a solver based on a first-order Godunov scheme with the HLL Riemann solver to solve SVE~\parencite{linLowcostDataefficientOndevice2025,chenAugmentedHLLRiemann2021,leonGodunovTypeSolutionsTransient2006}.
This numerical scheme is capable of dealing with abrupt variations and complex geometry with high accuracy~\parencite{chenAugmentedHLLRiemann2021,leonGodunovTypeSolutionsTransient2006,yuNewFormSaintVenant2020}.
The spatiotemporal domain is discretized with \(\Delta x = \qty{1}{\metre}\) and a Courant number (CFL) of 0.8 (\ref{sec:si:numerical-solver}).

\subsection{Leakage localization and quantification}\label{sec:leakage-localization-and-quantification}

To locate leaking pipes, we define a dimensionless pipe leakage evidence score (\(s_{\mathrm{leak}}\)) as
\begin{equation}
  s_{\mathrm{leak}} = \sqrt{\tilde{r}_{u}\tilde{r}_{d}}
  \label{eq:leakage-evidence-score}
\end{equation}
where \(\tilde{r}_{u}\) and \(\tilde{r}_{d}\) are the normalized mass balance residuals at the upstream and downstream manholes of the pipe, respectively.
For a manhole \(i\), the normalized mass balance residual \(\tilde{r}_{i}\) is calculated as
\begin{equation}
  \tilde{r}_{i}
  = \frac{\frac{1}{T}\sum_{t = 1}^{T}\left| r_{i}^{t} \right|}
    {\max_{j \in \mathcal{A}_{i}}\left| \mathcal{Q}_{j} \right|},
  \qquad r_{i}^{t} = \sum_{k}Q_{ik}^{t}.
  \label{eq:normalized-mass-balance-residual}
\end{equation}
Here, \(T\) is the number of time steps; \(r_{i}^{t}\) is the mass balance residual at manhole \(i\) at time \(t\); \(\mathcal{A}_{i}\) is the set of pipes connected to manhole \(i\); \(\mathcal{Q}_{j}\) is the full-pipe flow rate of pipe \(j\) calculated using Manning's formula (\autoref{eq:si:evaluation-full-pipe-flow}) for score normalization; \(Q_{ik}^{t} > 0\) denotes flow entering manhole \(i\), including flows from upstream pipes and lateral inflows.

For a pipe with leakage, we could accurately calculate the leakage volume \(V_{\mathrm{leak}}\) using the law of conservation of mass:
\begin{equation}
  \label{eq:leakage-volume}
  V_{\mathrm{leak}}
  = \int_{t_{1}}^{t_{2}}\left[ \hat{Q}_{\mathrm{out}}(t) - \hat{Q}_{\mathrm{in}}(t) \right] dt
  + \Delta V_{\mathrm{pipe}}
  \approx \sum_{t = 1}^{T}\left( \hat{Q}_{\mathrm{out}}^{t} - \hat{Q}_{\mathrm{in}}^{t} \right)\Delta t
\end{equation}
where \(\hat{Q}_{\mathrm{in}}(t)\) and \(\hat{Q}_{\mathrm{out}}(t)\) are the instantaneous flow rates into and out of the target pipe, respectively, calculated using the learned operator.
Although \(\Delta V_{\mathrm{pipe}}\) cannot be accurately calculated in leakage scenarios, its effect on \(V_{\mathrm{leak}}\) can be ignored with a sufficiently long horizon \(T\).

\subsection{Evaluation cases and study sites}\label{sec:evaluation-cases-and-study-sites}

Although the learned operator was trained exclusively on synthetic data driven by simple sinusoidal inflows, we evaluated it at both pipe and network scales using both direct field measurements and high-fidelity simulations driven by field-derived boundary conditions.
The evaluation involved three monitored real-life pipes, one synthetic pipe, and two real-life sewer networks.

The four pipe-level test cases have hydraulic parameter combinations and boundary conditions that do not appear in the training set, as shown in \autoref{tab:test-pipe-parameters}.
Pipes \#1-\#3 are real-world monitored pipes with field-measured water levels and flow rates, for which the roughness coefficients were determined through calibration.
Pipe \#4 is a synthetic pipe with prescribed geometric and hydraulic parameters.
Together, these cases span distinct hydraulic regimes: Pipe \#1 shows typical dry- and wet-weather sewer flow, Pipe \#2 features pump-driven high-frequency variation, Pipe \#3 exhibits wet-weather backwater effects, and Pipe \#4 represents severe surcharge conditions (\ref{sec:si:test-pipes}).

\begin{table}[htbp]
  \centering
  \singlespacing
  \caption{Geometric and hydraulic parameters of the test pipes}
  \label{tab:test-pipe-parameters}
  \small
  \begin{tabular}{lrrrr}
    \toprule
     & Diameter (\unit{\milli\metre}) & Length (\unit{\metre}) & Slope & Roughness\\
    \midrule
    Pipe \#1 & 1200 & 98 & 0.004 & 0.037\\
    Pipe \#2 & 600 & 64 & 0.006 & 0.018\\
    Pipe \#3 & 600 & 70 & 0.005 & 0.025\\
    Pipe \#4 & 1800 & 120 & 0.004 & 0.016\\
    \bottomrule
  \end{tabular}
\end{table}

We further evaluated the proposed method using a sewer network in Hong Kong.
The network is located in Kowloon City District, covers an area of \qty{1.2}{\kilo\metre\squared}, and serves a population of approximately 50,000~\parencite{zhuangAssessingRunoffControl2023}.
The sewer model after preprocessing contains 485 pipes with a total length of \qty{14.44}{\kilo\metre} and diameters from \qtyrange{150}{1800}{\milli\metre}.
We selected a rainfall event recorded by a nearby rain gauge from 19:00 to 20:15 on 28 April 2025.
The rainfall-runoff process was simulated using the Storm Water Management Model (SWMM) to generate nodal inflows, which were then used as boundary inputs for evaluating the proposed surrogate for network-scale hydraulic simulation (\ref{sec:si:test-sewer-system}).

Another sewer network located in the Greater Bay Area, China, was used to demonstrate the potential of the proposed method for leakage detection.
The sewer model after preprocessing contains 397 pipes with a total length of \qty{11.17}{\kilo\metre} and diameters from \qtyrange{150}{1800}{\milli\metre}.
We selected a typical dry weather event with a duration of \qty{24}{\hour} from real-life measurements.
Sewer flows were scaled based on the selected pattern and were fed into the sewer network via the upstream-most manholes.

    \section*{Data and code availability}\label{sec:data-and-code-availability}

The non-confidential code and data are available in our repository (\url{https://github.com/ruozhoulin/Generalized-Neural-operator-for-Sewer-Hydraulics}).

    \section*{Declaration of interests}\label{sec:declaration-of-interests}

The authors declare that they have no known competing financial interests or personal relationships that could have appeared to influence the work reported in this paper.

    \section*{Acknowledgements}\label{sec:acknowledgements}

The work described in this paper was supported by a grant from the Research Grants Council of the Hong Kong Special Administrative Region, China (Project No. CityU 11210825) and the National Natural Science Foundation of China (NSFC) (grant numbers 52400113).
Zhiguo Yuan is a Global STEM Professor jointly funded by the Innovation, Technology and Industry Bureau (“ITIB”) and Education Bureau (“EDB”) of the Government of the Hong Kong Special Administrative Region, China and acknowledges financial support from the Hong Kong Jockey Club for the JC STEM Lab of Sustainable Urban Water Management.

    \printbibliography[title={References}]
  \end{refsection}
  \clearpage

  \setcounter{section}{0}
  \setcounter{equation}{0}
  \setcounter{figure}{0}
  \setcounter{table}{0}
  \renewcommand{\thesection}{Supplementary Note \arabic{section}}
  \renewcommand{\theequation}{S\arabic{equation}}
  \renewcommand{\thefigure}{S\arabic{figure}}
  \renewcommand{\thetable}{S\arabic{table}}
  \renewcommand{\theHsection}{supplement.\arabic{section}}
  \renewcommand{\theHequation}{S\arabic{equation}}
  \renewcommand{\theHfigure}{S\arabic{figure}}
  \renewcommand{\theHtable}{S\arabic{table}}
  \begin{refsection}
    \section*{Supplementary Information}
\phantomsection
\addcontentsline{toc}{section}{Supplementary Information}
\label{sec:si:supplementary-information}

\section{Multi-branch neural operator}
\label{sec:si:multi-branch-operator}

The proposed multi-branch neural operator consists of 14 branches, one for each nominal pipe diameter; within each branch, the input vector \(\mathbf{x} \in \mathbb{R}^{d}\) is first mapped to a high-dimensional Fourier feature space through a Fourier feature embedding layer~\parencite{tancikFourierFeaturesLet2020,liModelingTransientMixed2024}:
\begin{equation}
  \gamma\left( \mathbf{x} \right) = \left\lbrack \sin\left( 2\pi\mathbf{Bx} \right),\cos\left( 2\pi\mathbf{Bx} \right) \right\rbrack,
  \label{eq:si:fourier-feature-embedding}
\end{equation}
where \(\mathbf{B} \in \mathbb{R}^{m \times d}\) is the trainable Fourier frequency matrix, initialized with entries sampled from a Gaussian distribution, \(B_{ij}\sim\mathcal{N}(0,\sigma^{2})\).
Here, \(d\) is the input dimension, \(m\) is the number of Fourier frequencies, and \(\sigma\) denotes the frequency scale.
The embedded feature vector \(\gamma(\mathbf{x}) \in \mathbb{R}^{2m}\) is then passed through a fully connected hidden layer, followed by layer normalization \(LN( \cdot )\) and a nonlinear activation \(\phi( \cdot )\).
The resulting hidden representation is subsequently mapped to the predicted output through a linear output layer:
\begin{equation}
  \begin{aligned}
    \mathbf{h} &= \phi\left(LN\left(\mathbf{W}_{1}\gamma(\mathbf{x}) + \mathbf{b}_{1}\right)\right),\\
    \hat{\mathbf{y}} &= \mathbf{W}_{2}\mathbf{h} + \mathbf{b}_{2}.
  \end{aligned}
  \label{eq:si:branch-network-prediction}
\end{equation}
where \(\mathbf{h} \in \mathbb{R}^{n_{h}}\) is the hidden representation, \(n_{h}\) is the number of hidden neurons, \(\mathbf{W}_{1} \in \mathbb{R}^{n_{h} \times 2m}\) and \(\mathbf{b}_{1} \in \mathbb{R}^{n_{h}}\) are the trainable parameters of the hidden layer, \(\mathbf{W}_{2} \in \mathbb{R}^{p \times n_{h}}\) and \(\mathbf{b}_{2} \in \mathbb{R}^{p}\) are the trainable parameters of the output layer, and \(\hat{\mathbf{y}} \in \mathbb{R}^{p}\) is the predicted output vector.
In this study, \(d = 16\) and \(p = 2\).
The hyperparameters \(m\), \(\sigma\), and \(n_{h}\) were determined using grid search, as detailed below.

We implemented the proposed model using PyTorch~\parencite{paszkePyTorchImperativeStyle2019}.
Trainable parameters were initialized using the Kaiming initialization~\parencite{heDelvingDeepRectifiers2015} and updated by the Adam optimizer~\parencite{kingmaAdamMethodStochastic2017} with a batch size of 8,192.
The model was trained by minimizing the mean squared error (MSE) loss.
The learning rate was initialized at \(10^{- 3}\) and automatically reduced by 50\% when the validation loss plateaued, using the ReduceLROnPlateau scheduler in PyTorch.
We set the maximum number of training epochs to 5,000.
To reduce the risk of overfitting, we applied early stopping when the validation performance ceased to improve.
Across our experiments, most training runs stopped before the 900th epoch.

\begin{figure}[htbp]
  \centering
  \includegraphics[width=\linewidth,height=0.62\textheight,keepaspectratio]{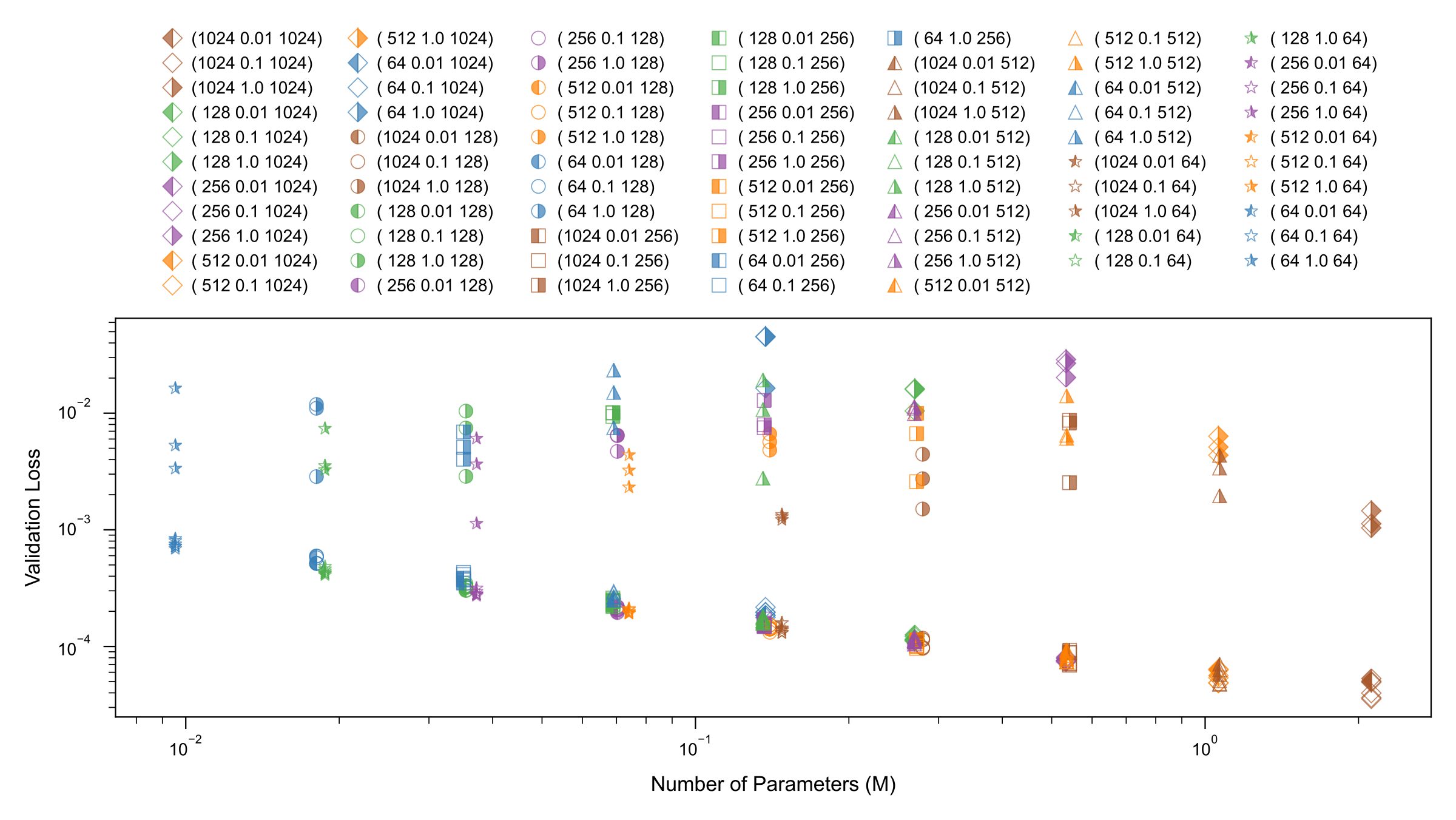}
  \caption{Performance of models with different hyperparameters on the validation set for \(D = 600\) mm.
Each tuple in the legend denotes (\(m\), \(\sigma\), and \(n_{h}\)), corresponding to the number of Fourier frequencies, the frequency scale, and the number of hidden neurons, respectively.}
  \label{fig:si:hyperparameter-validation}
\end{figure}

We used grid search to determine each branch's hyperparameters (\(m\), \(\sigma\), and \(n_{h}\)).
Taking the branch corresponding to \(D = 600\) mm as an example, we chose \(m = 256\), \(\sigma = 0.01\), and \(n_{h} = 512\) based on the grid-search results shown in \autoref{fig:si:hyperparameter-validation}, yielding 0.27M parameters as a tradeoff between accuracy and model complexity.
Other branches were selected in the same manner, with their hyperparameters and trainable parameter counts listed in \autoref{tab:si:branch-hyperparameters}.
The trained multi-branch neural operator is available in our open-source repository (\url{https://github.com/ruozhoulin/Generalized-Neural-operator-for-Sewer-Hydraulics}).

\begin{table}[htbp]
  \centering
  \singlespacing
  \caption{Hyperparameters and trainable parameter count for each branch determined by grid search.}
  \label{tab:si:branch-hyperparameters}
  \small
  \begin{tabular}{@{}rrrrrr@{}}
    \toprule
    Branch & $D$ (mm) & $m$ & $\sigma$ & $n_h$ & Trainable parameters\\
    \midrule
    1 & 150 & 512 & 0.1 & 256 & 271,618\\
    2 & 225 & 512 & 0.1 & 256 & 271,618\\
    3 & 300 & 512 & 0.1 & 256 & 271,618\\
    4 & 375 & 512 & 0.1 & 256 & 271,618\\
    5 & 450 & 512 & 0.1 & 256 & 271,618\\
    6 & 600 & 256 & 0.01 & 512 & 268,802\\
    7 & 750 & 512 & 0.1 & 256 & 271,618\\
    8 & 900 & 512 & 0.1 & 256 & 271,618\\
    9 & 1,050 & 512 & 0.1 & 256 & 271,618\\
    10 & 1,200 & 512 & 0.1 & 512 & 535,042\\
    11 & 1,350 & 512 & 0.1 & 256 & 271,618\\
    12 & 1,500 & 256 & 0.1 & 512 & 268,802\\
    13 & 1,650 & 512 & 0.1 & 256 & 271,618\\
    14 & 1,800 & 512 & 0.1 & 128 & 139,906\\
    \bottomrule
  \end{tabular}
\end{table}

\clearpage
\section{Numerical solver for Saint-Venant equations}
\label{sec:si:numerical-solver}

We solved the 1D Saint-Venant equations using a first-order Godunov scheme with the HLL Riemann solver.
The Saint-Venant equations can be written as
\begin{equation}
  \frac{\partial\mathbf{U}(x,t)}{\partial t}
  + \frac{\partial\mathbf{F}\left( \mathbf{U} \right)}{\partial x} = \mathbf{S}
  \label{eq:si:conservative-saint-venant}
\end{equation}
\begin{equation}
  \mathbf{U}\left(x,t\right) = \binom{A\ }{Q}
  \quad \mathbf{F}\left( \mathbf{U} \right) = \binom{Q}{\frac{Q^{2}}{A} + \frac{A\overline{P}}{\rho}}
  \quad \mathbf{S} = \binom{0}{gA{(S}_{0} - S_{f})}
  \label{eq:si:conservative-state-flux-source}
\end{equation}
where \(x\) denotes distance (\unit{\metre}); \(t\) denotes time (\unit{\second}); \(Q(x,t)\) denotes flow rate (\unit{\cubic\metre\per\second}); \(h(x,t)\) denotes water depth (\unit{\metre}); \(u(x,t)\) denotes velocity (\unit{\metre\per\second}); \(A(x,t)\) denotes cross-sectional flow area (\unit{\square\metre}); \(g\) denotes the gravitational acceleration (\unit{\metre\per\second\squared}); \(\overline{P}(x,t)\) denotes the average pressure of the water column over the cross-sectional area; \(\rho\) denotes liquid density; \(S_{0}\) denotes the pipe slope; \(S_{f}\) denotes the frictional slope evaluated using Manning's equation:
\begin{equation}
  S_{f} = \frac{n^{2}Q|Q|}{A^{2}R^{\frac{4}{3}}}
  \label{eq:si:solver-manning-friction-slope}
\end{equation}
where \(n\) is Manning's roughness coefficient and \(R\) is the hydraulic radius.

The first-order Godunov scheme discretizes the above equation as:
\begin{equation}
  \mathbf{U}_{i}^{n + 1} = \mathbf{U}_{i}^{n}
  - \frac{\Delta t}{\Delta x}\left( \mathbf{F}_{i + \frac{1}{2}}^{n} - \mathbf{F}_{i - \frac{1}{2}}^{n} \right)
  + \Delta tS_{i}^{n}
  \label{eq:si:godunov-update}
\end{equation}
where \(i\) is the spatial cell; \(n\) is the current time step; \(n + 1\) is the future time step; the time step \(\Delta t\) satisfies a CFL condition based on \(\max\left( |u| + \ c \right)\); \(\mathbf{F}_{i - \frac{1}{2}}^{n}\) and \(\mathbf{F}_{i + \frac{1}{2}}^{n}\) are the numerical interface fluxes obtained by solving Riemann problems using the HLL Riemann solver:
\begin{equation}
  \mathbf{F}_{i + \frac{1}{2}}^{n} = \left\{ \begin{matrix}
    \mathbf{F}_{L}^{n} & if\ s_{L} \geq 0 \\
    \frac{s_{R}\mathbf{F}_{L}^{n} - s_{L}\mathbf{F}_{R}^{n} + s_{R}s_{L}\left( \mathbf{U}_{R}^{n} - \mathbf{U}_{L}^{n} \right)}{s_{R} - s_{L}} & if\ s_{L} < 0 < s_{R} \\
    \mathbf{F}_{R}^{n} & if\ s_{R} \leq 0
  \end{matrix} \right.
  \label{eq:si:hll-interface-flux}
\end{equation}
\(\mathbf{F}_{L}^{n}\) and \(\mathbf{F}_{R}^{n}\) denote the fluxes in the cells immediately to the left and right of the interface, respectively; the wave-speed estimates were computed as \(s_{L} = \min\left( u_{L} - \ c_{L},\ u_{R} - \ c_{R} \right)\) and \(s_{R} = \max\left( u_{L} + \ c_{L},\ u_{R} + \ c_{R} \right)\), where \(c\) is the gravity-wave celerity.
Further details can be found elsewhere~\parencite{leonGodunovTypeSolutionsTransient2006}.

For a problem defined over \(x \in \lbrack 0,L\rbrack\) and \(t \in \lbrack T - \Delta t,T\rbrack\), the initial condition and the upstream and downstream boundary conditions must be specified, as discussed in \ref{sec:si:training-boundary-conditions}.

The Courant-Friedrichs-Lewy (CFL) number is a dimensionless parameter used to assess the numerical stability of explicit schemes for solving partial differential equations.
It represents the ratio of the distance travelled by the fastest hydraulic wave during one time step to the computational cell length.
For the one-dimensional Saint-Venant equations, it is expressed as
\begin{equation}
  CFL\  = \ \frac{(|u|\  + \ c)\Delta t}{\Delta x}
  \label{eq:si:courant-number}
\end{equation}
where \(u\) is the flow velocity, \(c\) is the gravity-wave celerity, \(\Delta t\) is the time-step size, and \(\Delta x\) is the spatial grid size.
For numerical stability, the CFL number is generally constrained to be below unity.
We chose \(CFL = 0.8\) in this study.

\clearpage
\section{Combinations of hydraulic parameters}
\label{sec:si:hydraulic-parameter-combinations}

For hydraulic simulation, a circular pipe is characterized by four geometric and hydraulic parameters: diameter (\(D\)), length (\(L\)), slope (\(S\)), and Manning's roughness coefficient (\(n\)).
The selection of each parameter is constrained by practical conditions, design codes, and engineering handbooks.

For a given region, the option for diameter is limited and called the nominal diameter.
For example, in Hong Kong, the nominal diameters are \(\{ 150,\allowbreak 225,\allowbreak 300,\allowbreak 375,\allowbreak 450,\allowbreak 600,\allowbreak 750,\allowbreak 900,\allowbreak 1050,\allowbreak 1200,\allowbreak 1350,\allowbreak 1500,\allowbreak 1650,\allowbreak 1800 \}\) mm.
Different regions may have slightly different nominal diameters.
In this study, we considered all nominal diameters in Hong Kong, and the proposed method can be easily extended to other nominal diameters by adding a corresponding branch for each diameter.
In Hong Kong, the maximum allowable pipe length varies with pipe diameter, as summarized in \autoref{tab:si:maximum-pipe-length}.
In this study, we considered pipes longer than \qty{10}{\metre} to exclude very short pipe segments that are less representative of the main sewer network.

\begin{table}[htbp]
  \centering
  \singlespacing
  \caption{Maximum allowable pipe length for different pipe diameters.}
  \label{tab:si:maximum-pipe-length}
  \small
  \begin{tabular}{@{}lr@{}}
    \toprule
    Diameter (\unit{\milli\metre}) & Maximum length (\unit{\metre})\\
    \midrule
    \(\leq 675\) & 80\\
    \(>675\) and \(\leq 1050\) & 100\\
    \(>1050\) & 120\\
    \bottomrule
  \end{tabular}
\end{table}

Regarding Manning's roughness coefficient, we considered values ranging from 0.01 to 0.04, as suggested by the SWMM user manual.
The pipe slope was constrained to be greater than 0.003, consistent with common drainage design requirements, and less than 0.02 to satisfy the mild-slope assumption underlying the Saint-Venant equations.
Although steeper pipes may occur in real sewer systems, their flow regimes deviate from the Saint-Venant equations and were therefore excluded from this study.
Each interval defined by the upper and lower bounds was sampled at five to seven selected points.

In addition, the design velocity for each pipe was constrained between 0.75 and \qty{3.0}{\metre}/s.
The lower bound was imposed to reduce the risk of sediment deposition, whereas the upper bound was used to limit potential pipe-wall erosion and abrasion.
The design velocity was calculated using Manning's formula as follows:
\begin{equation}
  u = \frac{1}{n}R^{\frac{2}{3}}\sqrt{S}
  \label{eq:si:manning-design-velocity}
\end{equation}
where \(u\) is the design velocity (\unit{\metre\per\second}); \(n\) is Manning's roughness coefficient; \(R\) is the hydraulic radius under full-flow conditions (\unit{\metre}); and \(S\) is the pipe slope (m/m).

In summary, the resulting dataset includes 2,871 pipes covering diverse combinations of pipe diameter, length, slope, and roughness.

\clearpage
\section{A seven-segment pipe for effective data generation}
\label{sec:si:seven-segment-data-generation}

\begin{figure}[htbp]
  \centering
  \includegraphics[width=\linewidth,height=0.62\textheight,keepaspectratio]{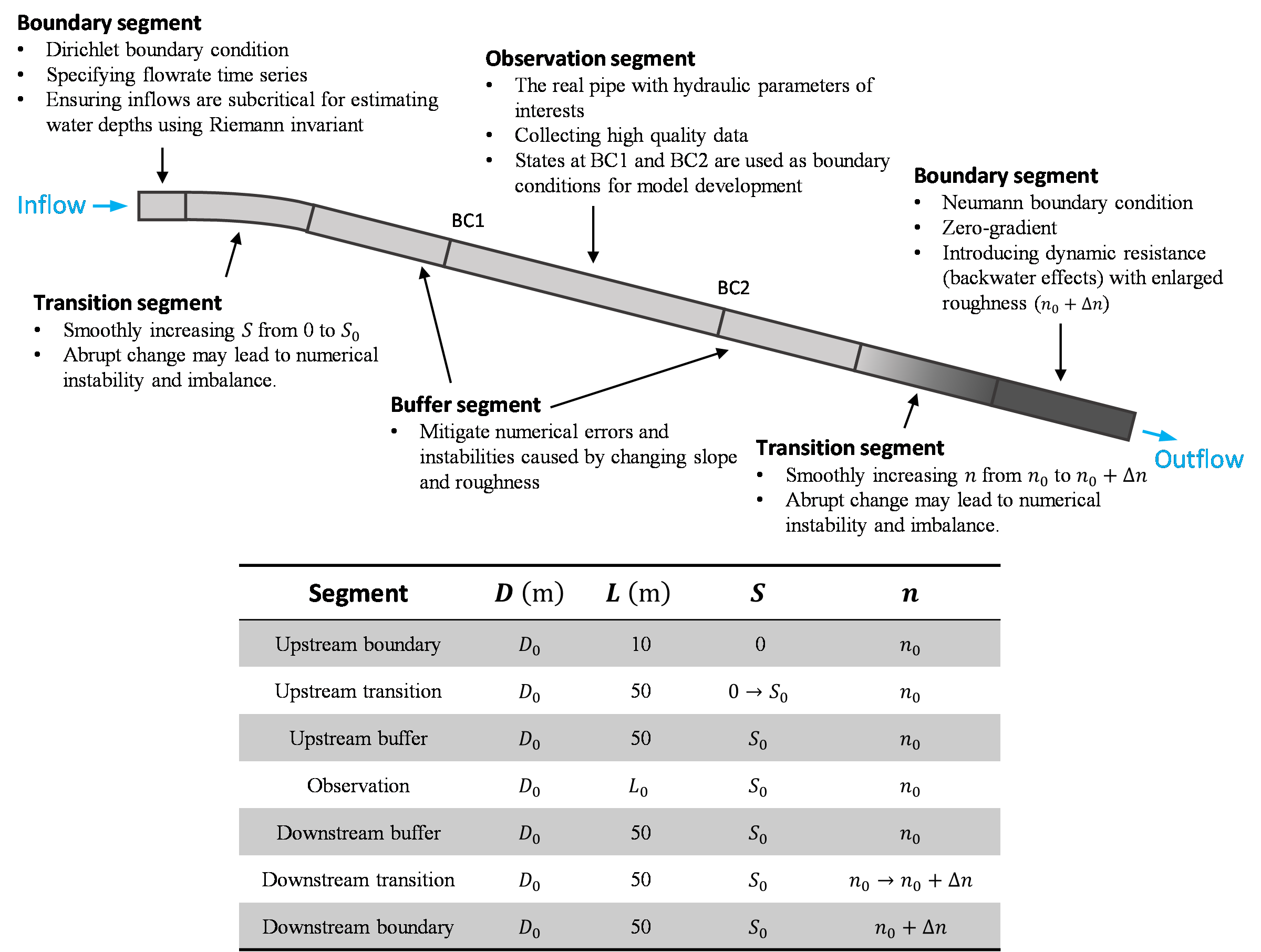}
  \caption{Schematic of the seven-segment computational domain for efficient data generation.}
  \label{fig:si:seven-segment-domain}
\end{figure}

Training a generalizable neural operator requires a comprehensive dataset covering a broad range of flow regimes.
To generate representative data efficiently, we embedded each pipe within a seven-segment computational domain, as shown in \autoref{fig:si:seven-segment-domain}.
During data generation, we only collected data from the observation segment for model development.
An inflow hydrograph (elaborated in \ref{sec:si:training-boundary-conditions}) was prescribed at the upstream boundary segment, which was assigned a zero slope to ensure subcritical flow regimes for handling boundary conditions using the Riemann invariant method.
In the subsequent upstream transition segment, we let the slope linearly increase from zero to the target slope (\(S_{0}\)) since abrupt changes in slope and roughness could cause numerical instability and imbalance.
Two buffer segments were placed around the observation segment to mitigate boundary-induced numerical artifacts.
The downstream transition segment linearly increased roughness from the target value (\(n_{0}\)) to \(n_{0} + \Delta n\), thereby introducing dynamic backwater effects.
This resistance is relatively weak, and may even be negligible, under low-flow conditions (e.g., dry weather flows), but becomes more pronounced as flow increases, as commonly observed during storm events when strong backwater effects and surcharge may occur.
Zero-gradient downstream boundary conditions were imposed for both water depth and flow rate to promote numerical stability.

\clearpage
\section{Boundary conditions for data generation}
\label{sec:si:training-boundary-conditions}

Although a comprehensive dataset would ideally include all possible inflow time series, such coverage is impractical.
Therefore, we generated representative inflow time series based on engineering judgment and prior domain knowledge.
Analysis of field measurements indicated that pipe inflows can be approximated as a superposition of sinusoidal waves, inspired by the Fourier representation.
Preliminary analysis has shown that the influence of prior flow conditions within a pipe is superseded by boundary conditions in less than \qty{5}{\minute}.
Accordingly, simple sinusoidal inflow profiles were used to generate representative sewer-flow scenarios.
Each flow rate series was defined as a sinusoidal function characterized by amplitude, frequency, phase, vertical offset, and duration.
Parameter combinations were selected based on prior knowledge and iterative trial runs to span a comprehensive set of scenarios.
An inflow time series for a pipe is defined as:
\begin{equation}
  Q(t) = Q_{m}\left\lbrack A\sin(\omega t + \phi) + k \right\rbrack
  \label{eq:si:sinusoidal-inflow}
\end{equation}
where \(Q_{m}\) is the full-pipe flow rate calculated using Manning's formula:
\begin{equation}
  Q_{m} = \frac{1}{n}AR^{\frac{2}{3}}\sqrt{S}
  \label{eq:si:training-full-pipe-flow}
\end{equation}

\begin{figure}[htbp]
  \centering
  \includegraphics[width=\linewidth,height=0.62\textheight,keepaspectratio]{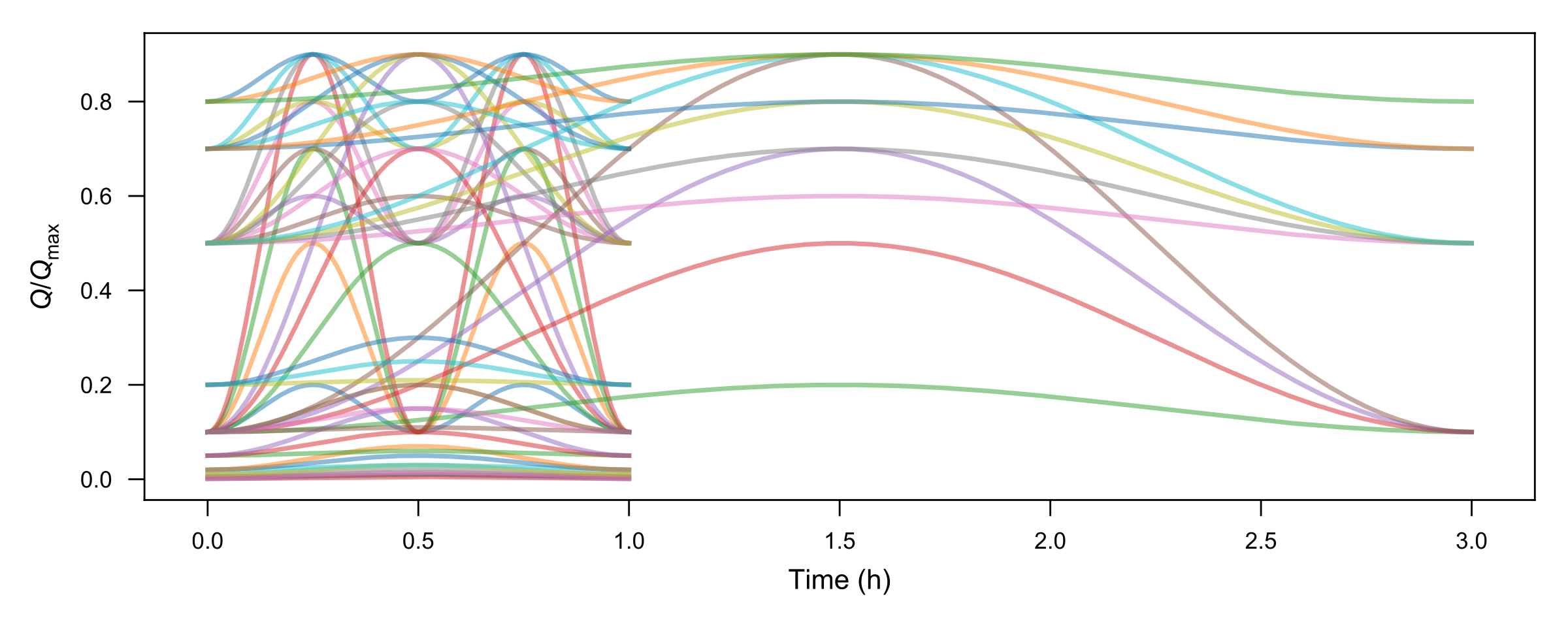}
  \caption{Overview of normalized upstream boundary conditions.}
  \label{fig:si:normalized-inflow-boundaries}
\end{figure}

We presented the normalized upstream boundary conditions used in this study in \autoref{fig:si:normalized-inflow-boundaries}.

All downstream boundary conditions were specified as zero-gradient conditions to improve numerical stability.
Downstream resistance (backwater effects) was introduced by increasing the roughness coefficient of the last pipe segment (\autoref{fig:si:seven-segment-domain}).
For each pipe and each inflow time series, we used a binary search to explore roughness values between 0.01 and 5.0, ensuring that both free-surface and pressurized flow conditions were represented in the observation segment.
This procedure yielded a diverse and comprehensive dataset for model development.
The resulting dataset covers a wide range of flow regimes, including backwater effects, free-surface flows, pressurized flows, transcritical flows, and transitional flows.
In total, we generated nearly 0.7 million scenarios.

\clearpage
\section{Data preprocessing}
\label{sec:si:data-preprocessing}

For data generated by the SVE solver, we removed duplicate data points and randomly subsampled those with water depths below \(0.05D\) or near steady state, limiting their proportion to no more than 10\%, to reflect practitioners' greater interest in dynamic and higher-flow conditions.

For data collected from real-life sewer systems, we applied linear interpolation to fill missing data and smoothing to reduce sensor noise.
\begin{equation}
  y = y_{0} + \frac{y_{1} - y_{0}}{t_{1} - t_{0}}\left( t - t_{0} \right)
  \label{eq:si:linear-interpolation}
\end{equation}
where \(\left( t_{0},y_{0} \right)\) denotes the last available measurement (water depth or flow rate) before the missing observation; \(\left( t_{1},y_{1} \right)\) denotes the next available measurement (water depth or flow rate) after the missing observation; and \((t,y)\) denotes the interpolated value (water depth or flow rate) at the time of the missing observation.

Specifically, a centered moving average filter with an 11-min window size was applied.
The window size was determined by trial and error to stand out the patterns.
\begin{equation}
  \hat{y}_{i} = \frac{1}{11}\sum_{j = i - 5}^{i + 5}y_{j}
  \label{eq:si:moving-average-smoothing}
\end{equation}
where \(\hat{y}_{i}\) is the smoothed value at time \(i\).

After dataset preparation, we standardized the data using z-score normalization, which transformed each variable to have a mean of 0 and a standard deviation of 1.
\begin{equation}
  z = \frac{x - \mu}{\sigma}
  \label{eq:si:z-score-normalization}
\end{equation}
where \(x\) is the original value; \(\mu\) is the mean of the variable; and \(\sigma\) is the standard deviation of the variable.

\clearpage
\section{Test set: four pipes}
\label{sec:si:test-pipes}

The test set includes four pipes with very different hydraulic characteristics that have not been seen in the training and validation sets.
Pipes \#1-\#3 are real-life pipes in Shenzhen, China.
Water depths and flow rates are continuously measured at manholes using integrated flow meters.
The sensor measures water levels and flow rates based on the principle of static pressure and Doppler effect, respectively, recording at an interval of 1 minute.
Pipe \#4 is synthetic to evaluate model performance under severe surcharges.

\begin{figure}[htbp]
  \centering
  \includegraphics[width=\linewidth,height=0.62\textheight,keepaspectratio]{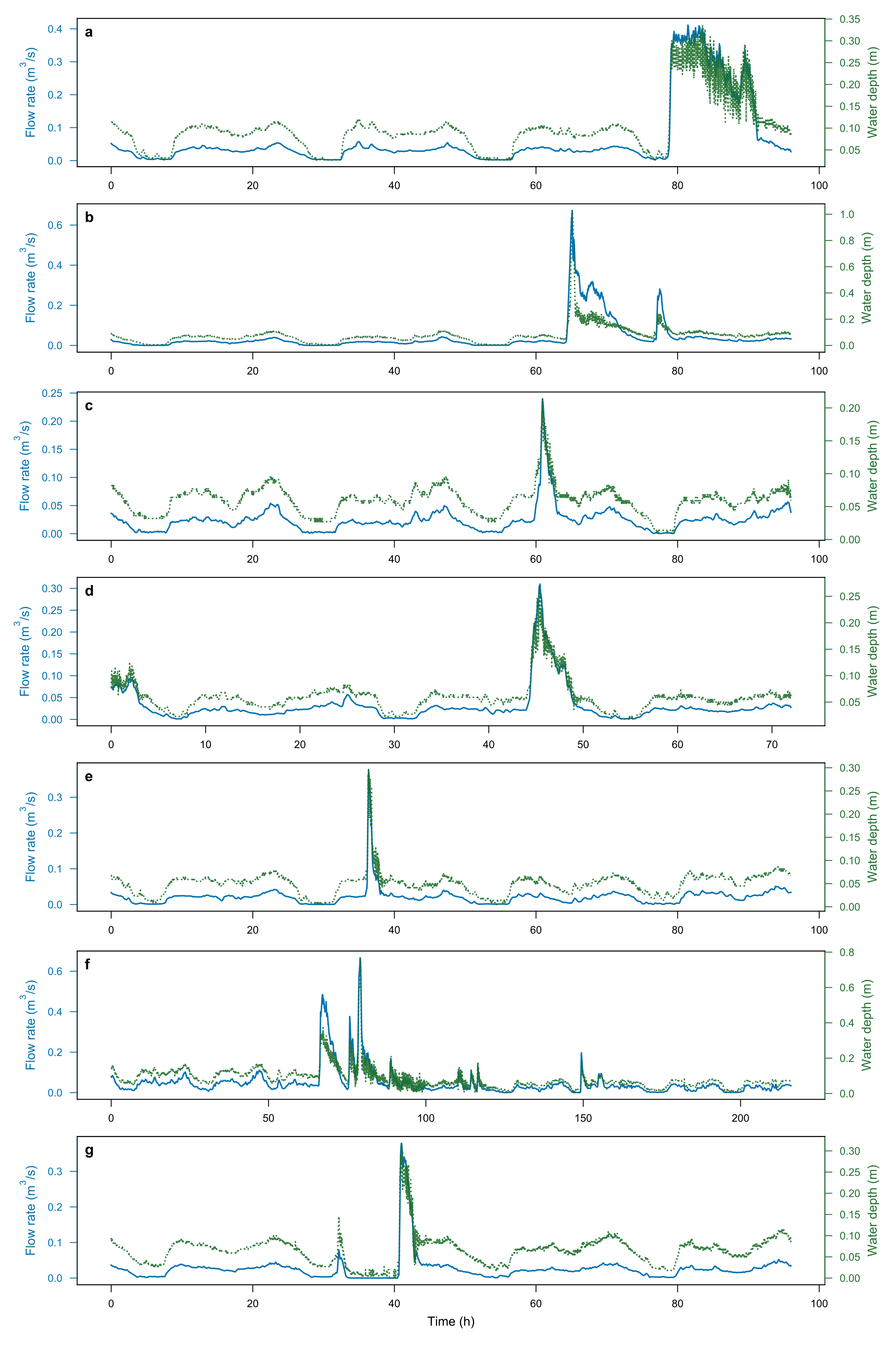}
  \caption{Measured flow rate and water depth for Pipe \#1 during seven wet weather events (a-g).
Blue solid lines indicate flow rate, and green dotted lines indicate water depth.}
  \label{fig:si:wet-weather-pipe-measurements}
\end{figure}

\textbf{Pipe \#1}.
This pipe has a diameter of \qty{1200}{\milli\metre}, a length of \qty{98}{\metre}, and a slope of 0.004, and was monitored at its upstream manhole, which is \qty{5.88}{\metre} deep.
The dataset covers \qty{672}{\hour} and includes both dry and wet weather flows, as shown in \autoref{fig:si:wet-weather-pipe-measurements}.
Downstream flows were not monitored.
To reconstruct flows inside this pipe for testing, we assumed \(n = 0.037\) (not in the training set), specified the measured inflows as the upstream boundary condition, and imposed a normal-flow condition downstream.

\begin{figure}[htbp]
  \centering
  \includegraphics[width=\linewidth,height=0.62\textheight,keepaspectratio]{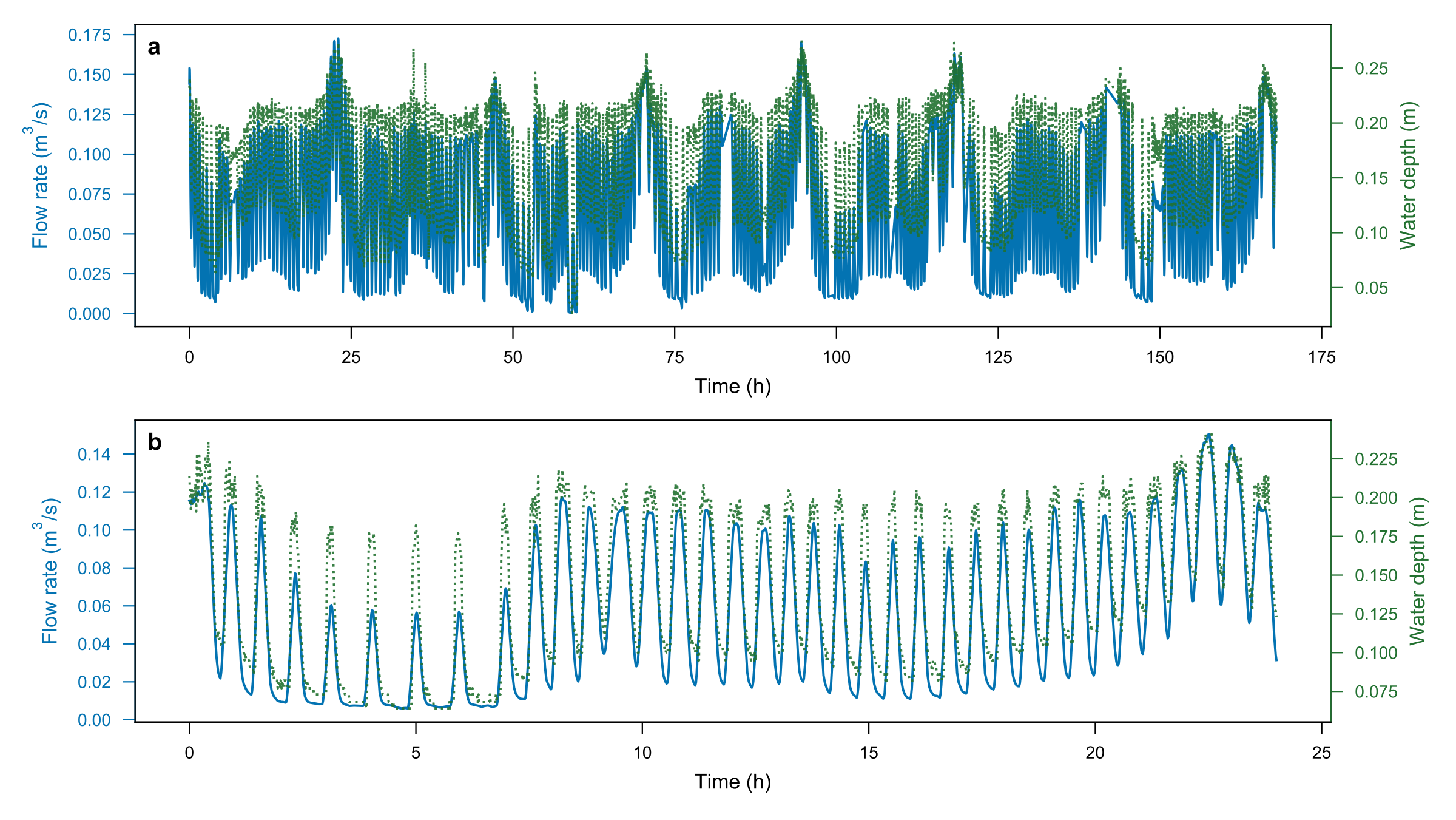}
  \caption{Measured flow rate and water depth for Pipe \#2 during two pump-driven events (a, b).
Blue solid lines indicate flow rate, and green dotted lines indicate water depth.}
  \label{fig:si:pump-driven-pipe-measurements}
\end{figure}

\textbf{Pipe \#2.} This pipe has a diameter of \qty{600}{\milli\metre}, a length of \qty{64}{\metre}, and a slope of 0.006, and was monitored at its upstream manhole, which is \qty{5.45}{\metre} deep and receives flows from an upstream pump station operating intermittently with a period of approximately \qty{40}{\minute}, as shown in \autoref{fig:si:pump-driven-pipe-measurements}.
The dataset covers \qty{192}{\hour}.
To reconstruct flows inside this pipe for testing, we assumed \(n = 0.018\) (not included in the training set), specified the measured inflows as the upstream boundary condition, and imposed a normal-flow condition downstream.

\begin{figure}[htbp]
  \centering
  \includegraphics[width=\linewidth,height=0.62\textheight,keepaspectratio]{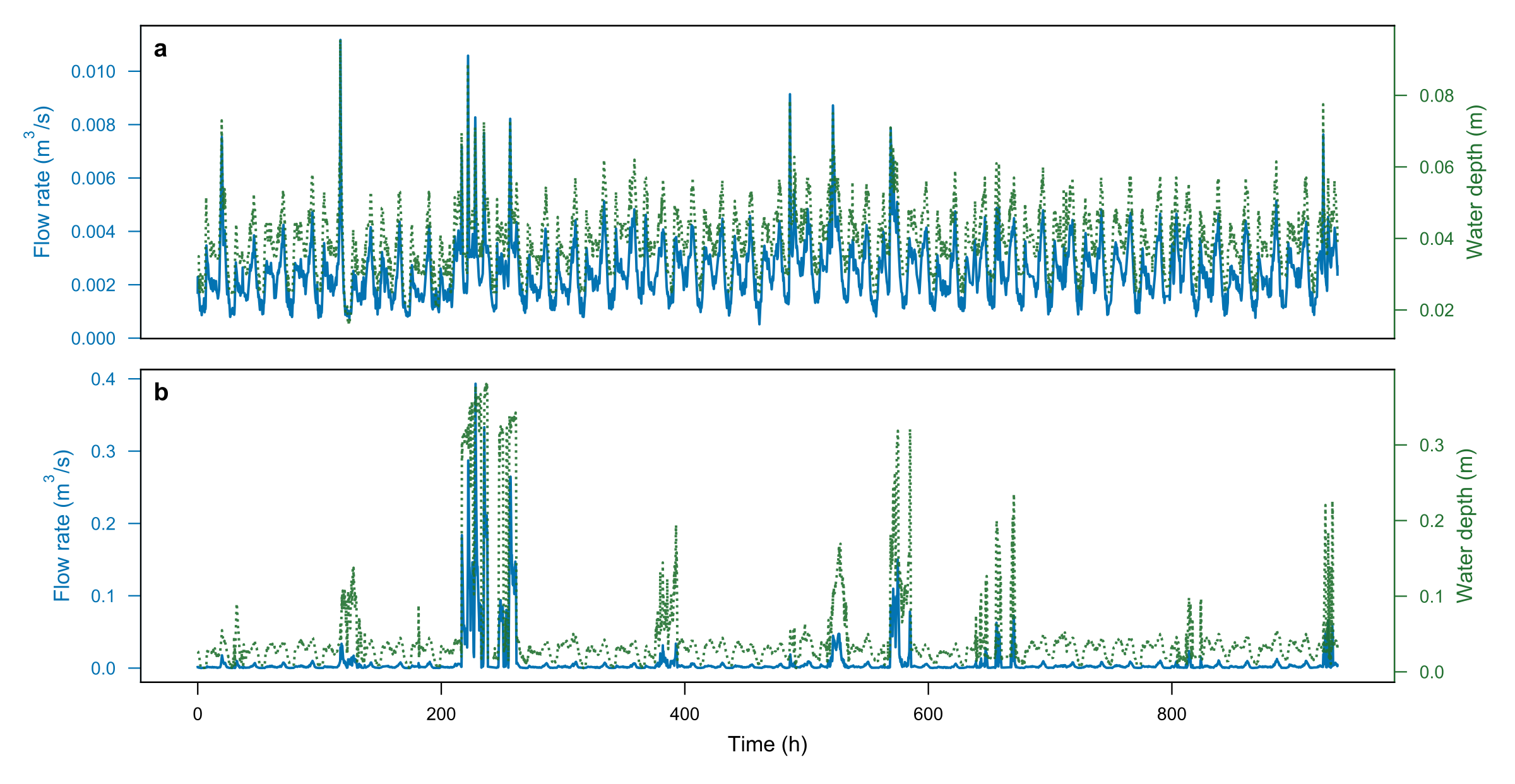}
  \caption{Measured flow rate and water depth for Pipe \#3 at the upstream manhole (a) and downstream manhole (b).
Blue solid lines indicate flow rate, and green dotted lines indicate water depth.}
  \label{fig:si:backwater-pipe-measurements}
\end{figure}

\textbf{Pipe \#3.} This pipe has a diameter of \qty{600}{\milli\metre}, a length of \qty{70}{\metre}, and a slope of 0.005, and was monitored at both ends.
The dataset includes both dry weather flows and wet weather flows with significant backwater effects caused by external inflows, as shown in \autoref{fig:si:backwater-pipe-measurements}.
The data covers 38 days with one record per minute.
Different from Pipe \#1 and \#2, the roughness coefficient (\(n = 0.025\)) of this pipe was determined by hydraulic calibration using the L-BFGS-B algorithm.
We reconstructed flows inside this pipe by specifying the measured inflow and downstream water depth as boundary conditions.

\begin{figure}[htbp]
  \centering
  \includegraphics[width=\linewidth,height=0.62\textheight,keepaspectratio]{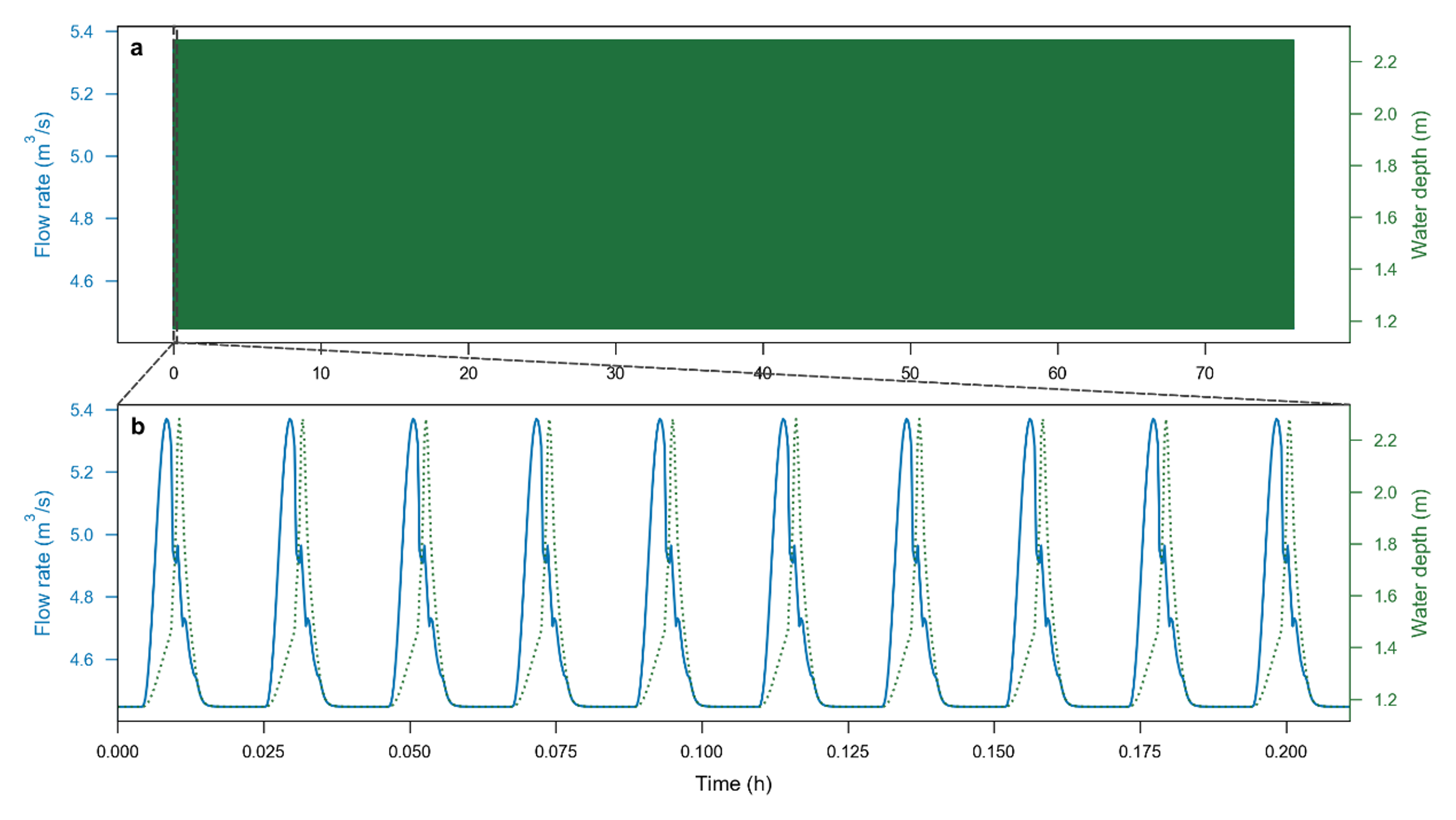}
  \caption{Synthetic flow rate and water depth for Pipe \#4 shown in (a) the full temporal domain and (b) a zoomed-in interval.
Blue solid lines indicate flow rate, and green dotted lines indicate water depth.}
  \label{fig:si:surcharged-pipe-simulation}
\end{figure}

\textbf{Pipe \#4.} This synthetic pipe has a diameter of \qty{1800}{\milli\metre}, a length of \qty{120}{\metre}, a slope of 0.004, and a roughness coefficient of 0.016, which do not exist in the training set.
We specified the inflow boundary condition upstream and a zero-gradient condition downstream.
The inflows caused severe surcharges most of the time, as shown in \autoref{fig:si:surcharged-pipe-simulation}.

In total, we generated around 381M data points with unseen features for testing.

\clearpage
\section{Test set: sewer system}
\label{sec:si:test-sewer-system}

The sewer network is located in Kowloon City District, Hong Kong, covering an area of \qty{1.2}{\kilo\metre\squared} and serving a population of around 50,000~\parencite{zhuangAssessingRunoffControl2023}.
The raw network data are available from the Common Spatial Data Infrastructure of Hong Kong (\url{https://portal.csdi.gov.hk/}).
We preprocessed the raw data to construct the sewer network for testing.

\textbf{Preprocessing}

First, we identified pipes with non-circular cross-sections and replaced them with circular pipes having the closest cross-sectional areas.
Pipes shorter than \qty{10}{\metre} were merged with adjacent pipes to avoid excessively short computational elements in network-level simulation.
In addition, two pipes had lengths that violated the design code; their lengths were adjusted to satisfy the code requirements.
Because the pipe slopes and roughness coefficients were unknown, we set them to 0.003 and 0.013, respectively, which are commonly used default values in practice.
We adopted the subcatchment delineation from a previous study~\parencite{zhuangAssessingRunoffControl2023}.

After preprocessing, the sewer model consisted of 485 circular pipes with diameters ranging from \qtyrange{150}{1800}{\milli\metre}, a total pipe length of \qty{14.44}{\kilo\metre}, and 5,917 subcatchments.

\textbf{Rainfall event}

We selected a rainfall event recorded at 15-min intervals from 19:00 to 20:15 on 28 April 2025 at a nearby rain gauge located at King's Park, approximately \qty{200}{\metre} from the sewer system.
The rainfall data were obtained from the Hong Kong Observatory (\url{https://data.weather.gov.hk/weatherAPI/}).
The corresponding rainfall hyetograph is shown in \autoref{tab:si:rainfall-hyetograph}.

\begin{table}[htbp]
  \centering
  \singlespacing
  \caption{Rainfall hyetograph for the selected rainfall event.}
  \label{tab:si:rainfall-hyetograph}
  \small
  \begin{tabular}{@{}cr@{}}
    \toprule
    Time (HH:mm) & Rainfall depth (mm)\\
    \midrule
    19:00 & 0\\
    19:15 & 1\\
    19:30 & 1\\
    19:45 & 2\\
    20:00 & 2\\
    20:15 & 0\\
    \bottomrule
  \end{tabular}
\end{table}

\textbf{Simulation}

We used SWMM to simulate the rainfall-runoff process with the semi-distributed nonlinear reservoir method and the Green-Ampt infiltration model.
The resulting runoff hydrographs were assigned to the sewer network as nodal lateral inflows.
Pipe flows were then routed using the high-fidelity SVE solver described in \ref{sec:si:numerical-solver}, with a spatial resolution of \(\Delta x = 1\) m.

\clearpage
\section{Evaluation metrics}
\label{sec:si:evaluation-metrics}

\textbf{Coefficient of determination (\(R^{2}\))}

\begin{equation}
  R^{2} = 1 - \frac{\sum_{i = 1}^{n}\left( y_{i} - \hat{y}_{i} \right)^{2}}{\sum_{i = 1}^{n}\left( y_{i} - \bar{y} \right)^{2}}
  \label{eq:si:coefficient-of-determination}
\end{equation}
where \(n\) is the number of data points; \(y_{i}\) is the observed value; \(\bar{y}\) is the average of observed values; and \(\hat{y}_{i}\) is the estimated value.

\textbf{Normalized mean absolute error (NMAE)}

This metric quantifies the error relative to the maximum hydraulic capacity of a pipe.
Given a pipe with a specified diameter (\(D\)), length (\(L\)), slope (\(S\)), and Manning's roughness coefficient (\(n\)), its NMAE for water depth is defined as:
\begin{equation}
  \frac{1}{n}\sum_{i = 1}^{n}\frac{\left| H_{i} - \hat{H}_{i} \right|}{D} \times 100\%
  \label{eq:si:water-depth-nmae}
\end{equation}
where \(H_{i}\) and \(\hat{H}_{i}\) are the \(i\)th water depth values from the high-fidelity SVE solver and the proposed operator, respectively.

Regarding flow rate, NMAE is defined as:
\begin{equation}
  \frac{1}{n}\sum_{i = 1}^{n}\frac{\left| Q_{i} - \hat{Q}_{i} \right|}{Q_{m}} \times 100\%
  \label{eq:si:flow-rate-nmae}
\end{equation}
where \(Q_{i}\) and \(\hat{Q}_{i}\) are the \(i\)th flow rate values from the high-fidelity SVE solver and the proposed operator, respectively.
\(Q_{m}\) is the full-pipe flow rate calculated using Manning's formula:
\begin{equation}
  Q_{m} = \frac{1}{n}AR^{\frac{2}{3}}\sqrt{S}
  \label{eq:si:evaluation-full-pipe-flow}
\end{equation}
where \(A\) and \(R\) are the cross-sectional area and hydraulic radius under full-flow conditions, respectively.

\clearpage

    \printbibliography[title={Supplementary References}]
  \end{refsection}
  \clearpage
\end{document}